\documentstyle[12pt,epsfig]{article}
\newcommand{\agt}{\,\rlap{\lower 3.5 pt \hbox{$\mathchar \sim$}} \raise 1pt
 \hbox {$>$}\,}
\newcommand{\alt}{\,\rlap{\lower 3.5 pt \hbox{$\mathchar \sim$}} \raise 1pt
 \hbox {$<$}\,}

\catcode`@=11
\newcount\@tempcntc
\def\@citex[#1]#2{\if@filesw\immediate\write\@auxout{\string\citation{#2}}\fi
  \@tempcnta\z@\@tempcntb\m@ne\def\@citea{}\@cite{\@for\@citeb:=#2\do
    {\@ifundefined
       {b@\@citeb}{\@citeo\@tempcntb\m@ne\@citea\def\@citea{,}{\bf ?}\@warning
       {Citation `\@citeb' on page \thepage \space undefined}}%
    {\setbox\z@\hbox{\global\@tempcntc0\csname b@\@citeb\endcsname\relax}%
     \ifnum\@tempcntc=\z@ \@citeo\@tempcntb\m@ne
       \@citea\def\@citea{,}\hbox{\csname b@\@citeb\endcsname}%
     \else
      \advance\@tempcntb\@ne
      \ifnum\@tempcntb=\@tempcntc
      \else\advance\@tempcntb\m@ne\@citeo
      \@tempcnta\@tempcntc\@tempcntb\@tempcntc\fi\fi}}\@citeo}{#1}}
\def\@citeo{\ifnum\@tempcnta>\@tempcntb\else\@citea\def\@citea{,}%
  \ifnum\@tempcnta=\@tempcntb\the\@tempcnta\else
   {\advance\@tempcnta\@ne\ifnum\@tempcnta=\@tempcntb \else \def\@citea{--}\fi
    \advance\@tempcnta\m@ne\the\@tempcnta\@citea\the\@tempcntb}\fi\fi}
\catcode`@=12
\begin{document}
\title{\vskip-3cm{\baselineskip14pt
\centerline{\normalsize DESY 99--005\hfill ISSN~0418--9833}
\centerline{\normalsize MPI/PhT/99--002\hfill}
\centerline{\normalsize hep--ph/9901348\hfill}
\centerline{\normalsize January 1999\hfill}}
\vskip1.5cm
Inclusive $J/\psi$ and $\psi(2S)$ Production from $B$ Decay in $p \bar p$
Collisions}
\author{B.A. Kniehl$^1$ and G. Kramer$^2$\\
$^1$ Max-Planck-Institut f\"ur Physik (Werner-Heisenberg-Institut),\\
F\"ohringer Ring 6, 80805 Munich, Germany\\
$^2$ II. Institut f\"ur Theoretische Physik\thanks{Supported by
Bundesministerium f\"ur Bildung, Wissenschaft, Forschung und Technologie
(BMBF), Bonn, Germany, under Contract 05~7~HH~92P~(0),
and by EU Fourth Framework Programme {\it Training and Mobility of
Researchers} through Network
{\it Quantum Chromodynamics and the Deep Structure of Elementary Particles}
under Contract FMRX--CT98--0194 (DG12 MIHT).},
Universit\"at Hamburg,\\
Luruper Chaussee 149, 22761 Hamburg, Germany}

\date{}
\maketitle
\begin{abstract}
Using information on $B$-meson fragmentation functions from CERN LEP~1 and 
adopting the nonrelativistic QCD factorization formalism proposed by Bodwin,
Braaten, and Lepage, we predict the transverse-momentum distribution of
$J/\psi$ mesons originating from the inclusive decays of $b$ hadrons
produced in $p\bar p$ collisions at the Fermilab Tevatron.
We determine the relevant colour-octet charmonium matrix elements from fits to 
CDF data on prompt charmonium hadroproduction and to CLEO data on charmonium 
production from $B$-meson decay.
Our predictions are found to agree well with recent CDF and D0 data.

\end{abstract}
\newpage 

\section{Introduction}

Recently the production of $J/\psi$ and $\psi^\prime$\footnote{We denote the
$\psi(2S)$ meson by $\psi^\prime$.} mesons in $p\bar p$ collisions at
$\sqrt s=1.8$~TeV was studied with the CDF detector at the Fermilab Tevatron
\cite{abe}.
The $J/\psi$ and $\psi^\prime$ mesons were reconstructed from their
$\mu^+\mu^-$ decay modes.
The inclusive production cross sections for both charmonium states were
measured as functions of the transverse momentum ($p_T$) in the central
region, corresponding to the rapidity ($y$) range $|y|<0.6$.
The CDF Collaboration was able to extract individual cross sections for
$J/\psi$ and $\psi^\prime$ mesons originating from weak decays of $b$ hadrons
and from prompt production.
In this way, rather accurate data for the inclusive $J/\psi$ and $\psi^\prime$
cross sections coming from the decays of $B$ mesons or other hadrons
containing $b$ quarks were obtained for $p_T$ values between 5 and 20~GeV.
These cross sections were compared with theoretical predictions based on
next-to-leading (NLO) calculations in quantum chromodynamics (QCD) with
massive $b$ quarks \cite{nas} and subsequent fragmentation of the $b$ quarks
into $b$ hadrons.
The $b$-hadron decays to $J/\psi+X$ and $\psi^\prime+X$ were described by a
parametrization of the momentum distribution measured by the CLEO
Collaboration \cite{bal}.
It was found that the data for $J/\psi$ ($\psi^\prime$) production were higher
than the QCD prediction by a factor of 2--4 (3--4) depending on the $p_T$ of
the $J/\psi$ ($\psi^\prime$) meson.
However, for larger $p_T$ values, {\it i.e.}\ $p_T\agt12$~GeV, the
experimental cross sections could be reproduced by the theoretical
calculations if the scale $\mu$, the $b$-quark mass $m_b$, and the parameter
$\epsilon$ in the Peterson \cite{pet} fragmentation function (FF) were
simultaneously reduced from $\mu=m_T=\sqrt{p_T^2+m_b^2}$ to $\mu=m_T/4$, from
$m_b=4.75$~GeV to $m_b=4.5$~GeV, and from $\epsilon = 0.006$ to
$\epsilon=0.004$, respectively.
Even with this choice of parameters, the measured cross section for $J/\psi$
production was still a factor of 2 above the prediction at $p_T\alt6$~GeV.
In the case of $\psi^\prime$ production, the cross section predicted with the
modified parameters is still below the data, although the discrepancy was only
about one standard deviation at large $p_T$.

For the nominal $J/\psi$ and $\psi^\prime$ predictions, it was assumed that the
nonperturbative part of the fragmentation of $b$ quarks into $b$ hadrons can
be described by a Peterson FF with $\epsilon=0.006$.
This value for $\epsilon$ was extracted more than ten years ago from a global
analysis of data on $B$ production in $e^+e^-$ annihilation at PETRA and PEP
\cite{chr}, based on Monte Carlo (MC) models which were in use at that time.
Due to the nonperturbative nature of the Peterson FF, the choice of $\epsilon$
must be backed up by other independent data, {\it e.g.}\ on $e^+e^-\to B+X$,
which must be analysed within the very theory that is used for the
interpretation of the CDF data, {\it i.e.}\ NLO QCD with massive quarks
($m_b\neq0$) and fixed order in $\alpha_s$ (massive scheme \cite{kni}).
In $e^+e^-$ annihilation, this theory is only reliable just above threshold,
where almost no data exist, except from the ARGUS and CLEO experiments at the
$\Upsilon(4S)$ resonance.
In this case, however, the $B$ mesons are only produced in pairs and not in
the fragmentation mode.
Therefore, the underlying description of $b\to B$ fragmentation in the massive
scheme, on which the comparison with the CDF data is based, is to a large
extent {\it ad hoc} and not supported by the analysis of independent data
within the same NLO perturbative scheme.
Reliable information on the FF for $b$ hadrons can only be gained from the
high-statistics experiments at CERN LEP~1.
At LEP~1, the production of $b$ quarks is enhanced as compared with the lower
$e^+e^-$ energies, so that the produced $b$ hadrons can be identified more
easily through their weak decays.
The fragmentation into $B$ mesons was measured by the OPAL Collaboration at
LEP~1 \cite{ale}.
Based on these data, we recently constructed FF's for $B$ mesons using three
different forms for the FF at the starting scale $\mu_0$, including the one by
Peterson {\it et al.}\ \cite{pet}, which yielded the lowest $\chi^2$ values at
leading order (LO) and NLO \cite{bin}.

In Ref.~\cite{bin}, the $b\to B$ FF was obtained using the so-called massless
scheme \cite{kni}.
In this scheme, the parton-level cross sections are calculated with $m_b=0$,
and the appearing collinear final-state singularities are factorized into the
FF's according to the modified minimal-subtraction ($\overline{\rm MS}$)
scheme.
A nonvanishing value for $m_b$ only appears in the initial conditions for the
FF's.
This scheme provides the appropriate approach for describing the fragmentation
of $b$ quarks into $B$ mesons at the $Z$-boson resonance, since the $b$ quarks
in the reaction $e^+e^-\to b\bar b\to B+X$ typically have large momenta.
A large-momentum $b$ quark essentially behaves like a massless particle,
radiating a large amount of its energy in the form of hard, collinear gluons.
This leads to the well-known logarithms of the form 
$\alpha_s\ln(M_Z^2/m_b^2)$ originating from collinear radiation in a scheme
where $m_b$ is taken to be finite.
These terms appear in all orders of perturbation theory.
The method for summing them is to introduce FF's and to absorb the
$m_b$-dependent logarithms into their evolution up to the factorization scale
of order $M_Z$.
If all terms of ${\cal O}(m_b^2/M_Z^2)$ are neglected and the
$\overline{\rm MS}$ subtraction scheme is adopted, then this approach is
equivalent to the massless scheme, where one puts $m_b=0$ from the beginning.
In the massless scheme, a nonperturbative FF is easily incorporated.
This allows one to transfer the information on the fragmentation process in
one reaction to other processes, {\it e.g.}\ $p\bar p\to B+X$.
This was done in Ref.~\cite{bin}, where the $b$-quark FF's obtained from fits to
the OPAL data were used to predict the differential cross section
$d\sigma/dp_T$ of $B$ production in $p\bar p$ scattering at $\sqrt s=1.8$~TeV.
These predictions were compared with data from the CDF Collaboration \cite{laa} 
and found to agree with them within errors.
In this paper, we use these results for $B$ production to predict the
distribution $d\sigma/dp_T$ of $J/\psi$ and $\psi^\prime$ mesons originating
from $b$-hadron decay.
For this purpose, we need a realistic description of the inclusive $B$ decays
into $J/\psi$ and $\psi^\prime$ mesons.
We adopt the parton-model description from Ref.~\cite{pal}, which allows one
to nicely interpret the CLEO data \cite{bal}.

While only $B^+$ and $B^0$ mesons and their antiparticles, which we 
collectively call $B$ mesons in the following, are produced at CLEO energies,
the CDF $J/\psi$ and $\psi^\prime$ samples also contain contributions from
$B_s$ mesons, $B_c$ mesons, $\Lambda_b$ baryons, etc.
We may safely ignore $B_c$ production, whose rate is expected to be about
$10^{-3}$ of the total $b$-hadron production rate \cite{bra}.
The $c$ quark in the initial state should enhance the branching ratio to 
charmonium, but $B_c$ production is expected to account for $\alt1\%$ of all
events of charmonium production from $b$-hadron decay at the Tevatron
\cite{cha}.
Unfortunately, there exist no measurements of the differential cross sections
of $B_s$ and $\Lambda_b$ production at LEP~1, which could be used to extract
FF's for these hadrons.
Moreover, there are no data on the momentum distributions of the $J/\psi$ and
$\psi^\prime$ mesons inclusively produced in $B_s$ and $\Lambda_b$ decays.
Only the branching fractions of $b\to\bar B_s$ and $b\to\Lambda_b$
(here $\Lambda_b$ stands for a collection of $b$ baryons) are known, which are
approximately 10\% each \cite{cas}.
Thus, the dominant channels for $b$-quark fragmentation are $b\to B^-$ and
$b\to\bar B^0$.
Due to the lack of detailed knowledge of the fragmentation into $B_s$ mesons
and $\Lambda_b$ baryons and their inclusive $J/\psi$ and $\psi^\prime$ decay
properties, we must estimate the $B_s$ and $\Lambda_b$ contributions using
information on $b\to B$ fragmentation and on the momentum distributions of the
inclusive $B\to J/\psi+X$ and $B\to\psi^\prime+X$ decays.
For simplicity, we assume that the $b\to\bar B_s$ and $b\to\Lambda_b$ FF's are
proportional to those of $b\to B^-$ \cite{bin}, and scale the latter by the
factor $B(b\to\bar B_s)/B(b\to B^-)$ and similarly for $b\to\Lambda_b$ 
adopting the values for these branching ratios from Ref.~\cite{cas}.
Furthermore, we approximate the momentum distributions of the $J/\psi$ 
($\psi^\prime$) mesons from the $B_s$ and $\Lambda_b$ decays by the one of
$B\to J/\psi+X$ ($B\to\psi^\prime+X$).

This paper is organized as follows.
In Sect.~2, we recall the framework for calculating the cross section of
$B$ production in $p\bar p$ collisions closely following our earlier work
\cite{bin}.
In Sect.~3, the momentum distributions of $J/\psi$, $\chi_{cJ}$, and 
$\psi^\prime$ mesons from $B$ decay, which enter the predictions of charmonium
production from $b$-hadron decay at the Tevatron, are described in the
framework of nonrelativistic QCD (NRQCD) \cite{bod}.
In Sect.~4, the relevant charmonium matrix elements are determined from CDF
\cite{abe} and CLEO \cite{bal} data on prompt charmonium production.
In Sect.~5, we calculate the $p_T$ spectra of $J/\psi$ and $\psi^\prime$ 
mesons originating from $b$-hadron decay at the Tevatron and compare them with
available CDF \cite{abe} and D0 \cite{abb} data.
Section~6 summarizes our conclusions.

\boldmath
\section{$B$-meson production in $p\bar p$ collisions}
\unboldmath

In Ref.~\cite{bin}, we presented LO and NLO predictions for the inclusive
cross section of $B$ production in $p\bar p$ collisions with $\sqrt s=1.8$~TeV
at the Tevatron.
This analysis provides the basis for the theoretical description of the
inclusive production of $J/\psi$ and $\psi^\prime$ mesons from $b$ hadrons at 
the Tevatron.
Having obtained the inclusive $B$ cross section $d\sigma/dp_T$, this cross
section is convoluted with the appropriately boosted longitudinal-momentum
distributions of the $J/\psi$ and $\psi^\prime$ mesons from $B$ decay, which 
are considered in the next section.
In this way, we obtain the bulk of the cross section, more than 80\% of it,
from which the additional contributions originating from the production of
$B_s$ mesons and $\Lambda_b$ baryons may be estimated.

Before we come to these points, we shortly recapitulate the input that was
assumed in Ref.~\cite{bin}.
The formalism used in Ref.~\cite{bin} is very similar to Ref.~\cite{bor}, where
inclusive light-meson production in $p\bar p$ collisions was studied in the
QCD-improved parton model.
We work at NLO in the $\overline{\rm MS}$ scheme with $N_f=5$ massless
flavors.
In this respect, we differ from any of the massive calculations \cite{nas},
where only $N_f=4$ active flavors are taken into account.
For the proton and antiproton parton density functions (PDF's) we use the set
CTEQ4M \cite{lai} with asymptotic scale parameter
$\Lambda_{\overline{\rm MS}}^{(5)}=202$~MeV or the more recent set MRST
\cite{mar} of the Durham-Oxford-Rutherford group with
$\Lambda_{\overline{\rm MS}}^{(5)}=211$~MeV, which corresponds to
$\Lambda_{\overline{\rm MS}}^{(4)}=300$~MeV and a $b$-quark mass of
$m_b=4.3$~GeV \cite{mar}.
The authors of Ref.~\cite{mar} presented new fits to the relevant
deep-inelastic-scattering data incorporating a more realistic description of
the heavy-quark PDF's, with an improved extrapolation from scales near
threshold to higher scales.
For the $B$ FF's, we adopt set NLO~P \cite{bin}, which uses the Peterson form
\cite{pet} for the $b\to B$ FF at the initial scale and yielded the best fits to
the OPAL data \cite{ale}.
The two alternative sets S and B, which are provided in Ref.~\cite{bin}, lead to
almost identical predictions for inclusive $B$ hadroproduction.
The evolution of the FF's is performed with
$\Lambda_{\overline{\rm MS}}^{(5)}=227$~MeV \cite{bin}.
The strong coupling constant $\alpha_s^{(5)}(\mu)$ is evaluated from the
two-loop formula adopting the $\Lambda_{\overline{\rm MS}}^{(5)}$ value from 
the selected set of proton PDF's.
We identify the factorization scales associated with the proton, antiproton,
and $B$ meson and collectively denote them by $M_f$.
We choose the renormalization and factorization scales to be
$\mu=M_f=2\xi m_T$, where $\xi$ is a number of order unity and
$m_T=\sqrt{p_T^2+m_\psi^2}$ is the transverse mass of the produced charmonium
state, which we generically denote by $\psi$.
Unless otherwise stated, we put $\xi=1$.
We recall that, in the case of inclusive $B$ production discussed in
Ref.~\cite{bin}, we instead took $m_T$ to be the $B$ transverse momentum.
In the present case, we consider it more appropriate to relate $m_T$ to the
final-state $\psi$ state.
The rationale is that we combine the $b\to B$ fragmentation and $B\to\psi+X$
decay processes into one single process.
This scale convention also considerably simplifies the computations, since the
evolution of the $B$ FF's are performed numerically in $x$ space \cite{jbi}.
Later on, we study the scale dependence of our results in order to get some
handle on the theoretical uncertainty related to this arbitrariness in
convention.
We adopt all kinematic conditions from Ref.~\cite{abe}.

When we present LO results, they are consistently computed using set CTEQ4L
\cite{lai} (MRSTLO \cite{mar}) of proton PDF's, set LO~P of $B$ FF's
\cite{bin}, the one-loop formula for $\alpha_s^{(5)}(\mu)$ with
$\Lambda^{(5)}=181$~MeV \cite{lai} ($\Lambda^{(5)}=132$~MeV \cite{mar}) 
and the LO hard-scattering cross sections.

\boldmath
\section{Charmonium production from $B$-meson decay}
\unboldmath

The inclusive decay $B\to J/\psi+X$ has three sources: the prompt production
of $J/\psi$ mesons and the two feed-down modes $B\to\psi^\prime+X$ followed by
$\psi^\prime\to J/\psi+X$ and $B\to\chi_{cJ}+X$, with $J=0,1,2$, followed by
$\chi_{cJ}\to J/\psi+\gamma$.
In the recent CLEO measurements \cite{bal}, these three sources were
disentangled and found to have the branching fractions $(0.80\pm0.08)\%$,
$(0.19\pm0.03)\%$, and $(0.13\pm0.02)\%$, respectively.
The total inclusive $J/\psi$ branching fraction was measured to be
$B(B\to J/\psi+X)=(1.12\pm0.07)\%$ \cite{bal}.
The inclusively produced $\psi^\prime$ mesons are all believed to be promptly
produced.
The CLEO Collaboration also measured the momentum distribution of the
$B\to J/\psi+X$ decay including all three sources and the one of the
$B\to \psi^\prime+X$ decay.
Furthermore, they presented the momentum distribution of the prompt
$B\to J/\psi+X$ decay, {\it i.e.}\ with the contributions due to the two
feed-down channels subtracted.

In order to predict the inclusive cross section of $J/\psi$ mesons from $B$
decay in $p\bar p$ collisions, we need the total $J/\psi$ momentum
distribution, {\it i.e.}\ the sum of the prompt and the two feed-down 
contributions.
So, the easiest way to incorporate the CLEO information would be to boost back
the CLEO data, measured in the CLEO laboratory frame, into the $B$ rest frame.
In this system, the distribution is assumed to be isotropic.
This distribution must then be boosted from the $B$ rest frame along the
momentum direction of the produced $B$ meson to the Tevatron laboratory frame.
This procedure can easily be implemented into an event generator to be used
for the analysis of the experimental data \cite{abe}.
Our procedure of incorporating the CLEO information is somewhat different.
We first calculate, within NRQCD \cite{bod}, the momentum distribution of
$J/\psi$ mesons promptly produced from $B$ decay in the $B$ rest system.
From this we derive the longitudinal-momentum distribution of the $J/\psi$
mesons for $B$ mesons in flight by performing an appropriate boost and
integrating over the transverse momentum components.
To adjust the input parameters, the momentum distribution in the $B$ rest
system is boosted to the CLEO laboratory frame and compared with the CLEO
data.
This already accounts for 71\% of the $J/\psi$ mesons coming from $B$ decay.
The same procedure can be applied to the $B\to\psi^\prime+X$ decay mode.
As we shall see in Sect.~4, in the case of $B\to\chi_{cJ}+X$, the NRQCD
prediction falls short of the CLEO data \cite{bal}, so that its normalization
must be adjusted by hand.
For simplicity, the feed-down from $\psi^\prime$ and $\chi_{cJ}$ mesons to
$J/\psi$ mesons is accounted for by including the respective branching 
fractions and assuming that the $J/\psi$ mesons receive the full momentum of
the primary charmonium states.

The momentum spectrum of the prompt, inclusive $B\to J/\psi+X$ decay has been
investigated just recently by Palmer, Paschos, and Soldan \cite{pal} in the
NRQCD approach \cite{bod}.
In this approach, the decay $b\to J/\psi+X$ is represented by a sum of
products, each of which consists of a short-distance coefficient for the
creation of a $c\bar c$ pair in a specific angular-momentum (${}^{2S+1}\!L_J$)
and color ($a=1,8$) configuration and a nonperturbative NRQCD matrix element
$\left\langle{\cal O}^{J/\psi}[\,\underline{a},{}^{2S+1}\!L_J]\right\rangle$,
which parametrizes the subsequent evolution of the intermediate 
$c\bar c[\,\underline{a},{}^{2S+1}\!L_J]$ state into the physical $J/\psi$
state (plus light hadrons) via the emission of soft gluons.
In their work, the transition $B\to b$ is described by two alternative models,
the parton model and the model based on Fermi momentum smearing \cite{ali}.
These models differ in the description of the long-distance transition
$B\to b$.
In the parton model approach, the transition $B\to b$ is parametrized by a
distribution function similar to the one used in deep-inelastic scattering.
In the Fermi motion approach, the bound-state corrections to the
free-$b$-quark decay are incorporated by giving the spectator quark a Fermi
motion inside the $B$ meson.
Since both models rely on unknown parameters, which are fixed by the CLEO data
on the $B\to J/\psi+X$ prompt, inclusive decay, we expect that the fine
details of describing the $B\to b$ transition are irrelevant for our purposes.
In particular, as we shall see, the momentum distribution of the $J/\psi$ 
mesons boosted to the Tevatron laboratory frame only depends on the bulk
properties of the momentum distribution in the $B$ rest system.
In the following, we adopt the parton-model description of $B\to b$.

In the perturbative NRQCD framework, one first calculates the decay rate
$\Gamma(b\to c\bar c[\,\underline{a},{}^{2S+1}\!L_J]+q)$ of a $b$ quark going
to a $c\bar c$ pair in state $[\,\underline{a},{}^{2S+1}\!L_J]$ plus a light 
quark $q$, where $q=d,s$.
Then, one obtains from this the $b\to J/\psi+X$ decay rate by exploiting the
NRQCD factorization theorem.
In the $b$-quark rest frame, the final formula reads \cite{pal}
\begin{equation}
\Gamma(b\to J/\psi+X)=\frac{G_F^2}{144\pi}|V_{cb}|^2m_cm_b^3
\left(1-\frac{4m_c^2}{m_b^2}\right)^2
\left[a\left(1+\frac{8m_c^2}{m_b^2}\right)+b\right],
\label{eq:par}
\end{equation}
where $G_F$ is Fermi's constant, $V_{cb}$ is an element of the 
Cabibbo-Kobayashi-Maskawa matrix, and $a$ and $b$ are NRQCD coefficients given
by
\begin{eqnarray}
a&=&(2C_+-C_-)^2
\frac{\left\langle{\cal O}^{J/\psi}[\,\underline{1},{}^3\!S_1]\right\rangle}
{3m_c^2} 
+(C_++C_-)^2
\nonumber\\
&&{}\times
\left[
\frac{\left\langle{\cal O}^{J/\psi}[\,\underline{8},{}^3\!S_1]\right\rangle}
{2m_c^2}
+\frac{\left\langle{\cal O}^{J/\psi}[\,\underline{8},{}^3\!P_1]\right\rangle}
{m_c^4}\right],
\nonumber\\
b&=&(C_++C_-)^2
\frac{\left\langle{\cal O}^{J/\psi}[\,\underline{8},{}^1\!S_0]\right\rangle}
{2m_c^2}.
\end{eqnarray}
Here, it is assumed that $m_q=0$ and $|V_{cs}|^2+|V_{cd}|^2=1$.
$C_{\pm}$ are the short-distance coefficients appearing in the
effective weak-interaction Hamiltonian that describes the $b\to c\bar cq$
transition.
We observe that the $b\to J/\psi+X$ decay rate essentially depends on two
parameters, $a$ and $b$, which are given in terms of the charm-quark mass 
$m_c$, the short-distance coefficients $C_{\pm}$, and the NRQCD color-singlet
and color-octet matrix elements.
In our numerical analysis, we use $G_F=1.16639\cdot10^{-5}$~GeV$^{-2}$,
$V_{cb}=0.0395$, $m_c=M_\psi/2$, where $M_\psi=3.09688$~GeV \cite{cas} is the
$J/\psi$ mass, $C_+=0.868$, and $C_-=1.329$ \cite{pal}.
The NRQCD matrix elements are specified in Sect.~4.

The inclusive decay of the $B$ meson is described by introducing the structure
function $f(x)$ for the transition $B\to b$.
It characterizes the distribution of the $b$-quark momentum inside the
$B$ meson and is usually parametrized by the Peterson form \cite{pet},
\begin{equation}
f(x)=N_{\epsilon}\frac{x(1-x)^2}{[(1-x)^2+\epsilon x]^2},
\end{equation}
where $\epsilon$ is a free parameter and $N_{\epsilon}$ is determined so 
that $f(x)$ is normalized to 1.
It is well known that the FF for the transition $b\to B$ peaks at large $x$.
Since we expect a similar behaviour for the inverse transition $B\to b$,
the Peterson ansatz should be well suited.
Specifically, we adopt the value $\epsilon=0.0126$, which was determined in
the LO analysis of Ref.~\cite{bin}.

Using the formalism of Ref.~\cite{pal}, we obtain the following formula for
the inclusive $B\to J/\psi+X$ decay width, differential in the $J/\psi$
three-momentum $\mbox{\boldmath$k_\psi^\prime$}$, in an arbitrary frame, where
the $B$ meson has three-momentum $\mbox{\boldmath$P_B$}$:
\begin{eqnarray}
E_BE_\psi^\prime\frac{d^3\Gamma^\prime}{d^3k_\psi^\prime}
&=&\frac{G_F^2M_\psi^3}{288\pi^2}|V_{cb}|^2
\left\{\left[a-b+(a+b)\frac{M_B^2}{2M_\psi^2}(x_++x_-)^2\right]
\left[f(x_+)+f(x_-)\right]\right. 
\nonumber\\
&&{}+\left.(a+b)\frac{M_B^2}{2M_\psi^2}(x_++x_-)(x_+-x_-)
\left[f(x_+)-f(x_-)\right]\right\},
\label{eq:cov}
\end{eqnarray}
where
$E_B=\sqrt{\mbox{\boldmath$P_B^{\phantom{2}}$}\!\!\!{}_{\phantom{\bf B}}^2
+M_B^2}$,
$E_\psi=\sqrt{\mbox{\boldmath$k_\psi^{\phantom{2}}$}\!\!\!
{}_{\phantom{\psi}}^2+M_\psi^2}$,
and
\begin{equation}
x_\pm=\frac{P_B\cdot k_\psi^\prime\pm
\sqrt{(P_B\cdot k_\psi^\prime)^2-M_B^2M_\psi^2}}{M_B^2}.
\end{equation}
Here, a prime is introduced to discriminate quantities referring to the system
with finite $\mbox{\boldmath$P_B$}$ from their counterparts in the $B$ rest
frame.
A similar formula applies to the inclusive $B\to\psi^\prime+X$ decay.

Equation~(\ref{eq:cov}) is used in two ways.
First, it is applied in the CLEO laboratory frame, where
$|$\mbox{\boldmath$P_B$}$|=\sqrt{M_\Upsilon^2/4-M_B^2}=0.341$~GeV, after
integration over the direction of $\mbox{\boldmath$k_\psi^\prime$}$, to
constrain the theoretical input entering $a$ and $b$ by comparison with the
CLEO data \cite{bal}.
For the prompt $B\to J/\psi+X$ decay, this was already done in Ref.~\cite{pal}.
A convenient alternative to doing the angular integration in the CLEO 
laboratory system is to carry it out in the $B$ rest frame and then to perform
a boost to the CLEO laboratory system.
In the $B$ rest frame, the decay width differential in
$k_\psi=|$\mbox{\boldmath$k_\psi$}$|$ is easily obtained as \cite{pal}
\begin{eqnarray}
\frac{d\Gamma}{dk_\psi}
&=&\frac{G_F^2M_\psi^3}{72\pi M_B}|V_{cb}|^2\frac{k_\psi^2}{E_\psi}
\left\{\left[a-b+(a+b)\frac{M_B^2}{2M_\psi^2}(x_++x_-)^2\right]
\left[f(x_+)+f(x_-)\right]\right. 
\nonumber\\
&&{}+\left.(a+b)\frac{M_B^2}{2M_\psi^2}(x_++x_-)(x_+-x_-)
\left[f(x_+)-f(x_-)\right]\right\},
\label{eq:pps}
\end{eqnarray}
where $x_\pm$ is now given by
\begin{equation}
x_\pm=\frac{E_\psi\pm k_\psi}{M_B}.
\end{equation}
Here, $k_\psi$ varies in the interval $0\le k_\psi\le k_{\psi,\rm max}$, where
\begin{equation}
k_{\psi,\rm max}=\frac{M_B^2-M_\psi^2}{2M_B}.
\label{eq:max}
\end{equation}
In the CLEO laboratory frame, we then have
\begin{equation}
\frac{d\Gamma}{dk_\psi^\prime}=\frac{M_Bk_\psi^\prime}{2P_BE_\psi^\prime}
\int_{k_1}^{k_2}\frac{dk_\psi}{k_\psi}\,\frac{d\Gamma}{dk_\psi},
\label{eq:res}
\end{equation}
where
\begin{equation}
k_1=\frac{|E_Bk_\psi^\prime-P_BE_\psi^\prime|}{M_B},\qquad
k_2=\min\left(\frac{E_Bk_\psi^\prime+P_BE_\psi^\prime}{M_B},k_{\psi,\rm max}
\right).
\end{equation}
Here, $k_\psi^\prime$ varies in the interval
$\max(k_-^\prime,0)\le k_\psi^\prime\le k_+^\prime$, where
\begin{equation}
k_\pm^\prime=\frac{P_B(M_B^2+M_\psi^2)\pm E_B(M_B^2-M_\psi^2)}{2M_B^2}.
\end{equation}
Notice that $d\Gamma$ on the left-hand side of Eq.~(\ref{eq:res}) refers to
the $B$ rest frame, so that integrating Eq.~(\ref{eq:res}) over
$k_\psi^\prime$ yields the proper $B\to J/\psi+X$ decay width.

The second application of Eq.~(\ref{eq:cov}) concerns the calculation of the
$B\to J/\psi+X$ decay distribution in the component $k_L^\prime$ of the
$J/\psi$ three-momentum that is parallel to $\mbox{\boldmath$P_B$}$.
For this purpose, we must integrate Eq.~(\ref{eq:cov}) over the orthogonal
momentum component $k_T^\prime$.
This can be directly done leading to
\begin{eqnarray}
\frac{d\Gamma}{dk_L^\prime}
&=&\frac{G_F^2M_\psi^3}{288\pi M_B}|V_{cb}|^2
\int_0^{k_{T,\rm max}^{\prime2}}\frac{dk_T^{\prime2}}{E_\psi^\prime}
\left\{\left[a-b+(a+b)\frac{M_B^2}{2M_\psi^2}(x_++x_-)^2\right]
\left[f(x_+)+f(x_-)\right]\right. 
\nonumber\\
&&{}+\left.(a+b)\frac{M_B^2}{2M_\psi^2}(x_++x_-)(x_+-x_-)
\left[f(x_+)-f(x_-)\right]\right\},
\label{eq:lon}
\end{eqnarray}
where
\begin{equation}
k_{T,\rm max}^{\prime2}=\left(\frac{M_B^2+M_\psi^2+2p_Bk_L}{2E_B}\right)^2
-M_\psi^2-k_L^{\prime2}
\end{equation}
and $k_-^\prime\le k_L^\prime\le k_+^\prime$.
For most of our applications, $P_B=|\mbox{\boldmath$P_B$}|\gg M_B$ is 
satisfied.
In this limit, an asymptotic formula for $d\Gamma/dx$, where
$x=k_L^\prime/P_B$, can be derived.
Defining $r=M_\psi^2/M_B^2$ and $t=k_T^{\prime2}/M_B^2$, we have
\begin{eqnarray}
\frac{d\Gamma}{dx}(x,P_B)&=&\frac{G_F^2M_B^3M_\psi}{288\pi x}|V_{cb}|^2 
\int_0^{(x-r)(1-x)}dt\left\{\left[r(a-b)+\frac{a+b}{2}(x_++x_-)^2\right]
\right.
\nonumber\\
&&{}\times\left.\left[f(x_+)+f(x_-)\right]
+\frac{a+b}{2}(x_+-x_-)^2\left[f(x_+)-f(x_-)\right]\right\},
\label{eq:inf}
\end{eqnarray}
where
\begin{equation}
x_\pm=\frac{1}{2}\left(x+\frac{r+t}{x}\pm
\sqrt{\left(x+\frac{r+t}{x}\right)^2-4r}\,\right),
\end{equation}
and $r\le x\le1$.
An alternative way of calculating the distribution in Eq.~(\ref{eq:lon}) is to
transform the $k_T^\prime$ integration into an integration over $k_\psi$ in
the $B$ rest frame, in a way similar to Eq.~(\ref{eq:res}).
This leads to
\begin{equation}
\frac{d\Gamma}{dk_L^\prime}=\frac{M_B}{2E_B}
\int_{k_{\psi,\rm min}}^{k_{\psi,\rm max}}\frac{dk_\psi}{k_\psi}\,
\frac{d\Gamma}{dk_\psi},
\label{eq:alt}
\end{equation}
where $d\Gamma/dk_\psi$ and $k_{\psi,\rm max}$ are given in
Eqs.~(\ref{eq:pps}) and (\ref{eq:max}), respectively, and
\begin{equation}
k_{\psi,\rm min}=\frac{|E_Bk_L^\prime-P_Bm_L^\prime|}{M_B},
\end{equation}
with $m_L^\prime=\sqrt{k_L^{\prime2}+M_\psi^2}$.

As already mentioned in Sect.~2, we incorporate the $B\to \psi+X$ decay in our
calculation of inclusive $B$ hadroproduction by defining effective FF's for
the transition of the partons $i$ that come out of the hard scattering to the
$\psi$ mesons as
\begin{equation}
D_{i\to\psi}(x,M_f^2)=\int_x^1dz\,D_{i\to B}\left(\frac{x}{z},M_f^2\right)
\frac{1}{\Gamma_B}\,\frac{d\Gamma}{dz}(z,P_B),
\label{eq:eff}
\end{equation}
where $D_{i\to B}(x,M_f^2)$ are the nonperturbative FF's determined in
Ref.~\cite{bin}, $\Gamma_B=1/\tau_B$ is the total $B$ decay width, and
$d\Gamma(x,P_B)/dx$ is obtained from Eqs.~(\ref{eq:lon}) or (\ref{eq:alt}).
For given $\psi$ transverse momentum $p_T$ and rapidity $y$, $P_B$ in
Eq.~(\ref{eq:eff}) is given by $P_B=\sqrt{p_T^2+m_T^2\sinh^2y}/z$, where
$m_T$ is the $\psi$ transverse mass.
For reasons explained above, we choose the factorization scale to be
$M_f=2\xi m_T$.
In our numerical analysis, we use the $B^+/B^0$ average values
$M_B=5.2791$~GeV and $\tau_B=1.61$~ps \cite{cas} for the $B$ mass and
lifetime, respectively.

In the remainder of this section, we discuss the decays $B\to\chi_{cJ}+X$,
with $J=0,1,2$, within the NRQCD framework.
They were not considered in Ref.~\cite{pal}.
They contribute via radiative feed-down to the yield of $J/\psi$ mesons from
$b$-hadron decay measured at the Tevatron and must, therefore, be included in
our analysis.
We find that Eq.~(\ref{eq:par}) and the subsequent equations derived from it
carry over to these cases if we substitute everywhere $\psi$ by $\chi_{cJ}$,
insert
\begin{eqnarray}
a&=&(C_++C_-)^2
\frac{\left\langle{\cal O}^{\chi_{c0}}
[\,\underline{8},{}^3\!S_1]\right\rangle}{2m_c^2},
\nonumber\\
a&=&(2C_+-C_-)^2\frac{2\left\langle{\cal O}^{\chi_{c0}}
[\,\underline{1},{}^3\!P_0]\right\rangle}{m_c^4} 
+(C_++C_-)^2\frac{3\left\langle{\cal O}^{\chi_{c0}}
[\,\underline{8},{}^3\!S_1]\right\rangle}{2m_c^2},
\nonumber\\
a&=&(C_++C_-)^2
\frac{5\left\langle{\cal O}^{\chi_{c0}}
[\,\underline{8},{}^3\!S_1]\right\rangle}{2m_c^2},
\end{eqnarray}
for $J=0,1,2$, respectively, and put $b=0$ in all three cases.

\section{Nonperturbative charmonium matrix elements}

In this section, we determine the leading NRQCD colour-octet matrix elements
of the $J/\psi$, $\psi^\prime$, and $\chi_{cJ}$ mesons from fits to the CDF
data of their prompt hadroproduction \cite{abe}, imposing the requirement that
the branching fractions of the inclusive $B$ decays into these charmonium 
states, calculated as described in Sect.~3, agree with the CLEO results
\cite{bal} whenever this is possible.
This complements our previous analysis \cite{bak}, where we determined the
$J/\psi$ and $\chi_{cJ}$ colour-octet matrix elements from the CDF data
\cite{abe} alone.

We start by repeating the $J/\psi$ analysis of Ref.~\cite{bak} for the
$\psi^\prime$ mesons.
We extract
$\left\langle{\cal O}^{\psi^\prime}[\,\underline{1},{}^3\!S_1]\right\rangle$
from the measured $\psi^\prime\to\mu^+\mu^-$ decay width \cite{cas} using the
QCD-improved formula (3) of Ref.~\cite{bak}.
We then determine the leading colour-octet matrix elements from the CDF data
sample of prompt $\psi^\prime$ hadroproduction \cite{abe}, which contains 5
data points.
Similarly to the $J/\psi$ analysis reported in Ref.~\cite{bak}, we obtain
$\left\langle{\cal O}^{\psi^\prime}[\,\underline{8},{}^3\!S_1]\right\rangle$
from the upper part of the $p_T$ spectrum (last 2 data points) working in the
fragmentation picture, where the $c\bar c$ bound state is created from a
single high-energy gluon, charm quark or antiquark which is close to its mass
shell.
In a second step, we extract the linear combination
\begin{equation}
M_r^{\psi^\prime}=
\left\langle{\cal O}^{\psi^\prime}[\,\underline{8},{}^1\!S_0]\right\rangle
+\frac{r}{m_c^2}
\left\langle{\cal O}^{\psi^\prime}[\,\underline{8},{}^3\!P_0]\right\rangle,
\label{eq:mr}
\end{equation}
from the lower part of the $p_T$ spectrum (first 4 data points) adopting the
fusion picture, where the $c\bar c$ bound state is formed within the primary
hard-scattering process. 
Here, $r$ is chosen in such a way that the superposition of these two channels
is insensitive to precisely how they are weighted relative to each other.
The results obtained with set CTEQ4L \cite{lai} of proton PDF's are listed in 
Table~\ref{t1} together with our previous LO results for $J/\psi$ mesons
\cite{bak}.
Our result for $M_r^{\psi^\prime}$ agrees very well with the value
$M_3^{\psi^\prime}=(1.8\pm0.6)\cdot10^{-2}$~GeV$^3$ found by Cho and Leibovich
\cite{cho} from a fit to earlier CDF data, while our result for
$\left\langle{\cal O}^{\psi^\prime}[\,\underline{8},{}^3\!S_1]\right\rangle$
is somewhat larger than their value
$\left\langle{\cal O}^{\psi^\prime}[\,\underline{8},{}^3\!S_1]\right\rangle=
(4.6\pm1.0)\cdot10^{-3}$~GeV$^3$.
Figure~\ref{f1} illustrates, for the fusion picture, how the theoretical cross
section of prompt $\psi^\prime$ hadroproduction compares with the 
corresponding CDF data \cite{abe} and how it is decomposed into its
colour-singlet and colour-octet components.
The total cross section in the fragmentation picture, which is complementary 
to the fusion picture, is not shown in Fig.~\ref{f1} because it would be
hardly distinguishable from the solid line.
It is almost entirely due to the $[\,\underline{8},{}^3\!S_1]$ channel, while
the $[\,\underline{1},{}^3\!S_1]$ contribution is approximately down by a
factor of 60.
The goodness of the fits in the fusion and fragmentation pictures is measured
in terms of the $\chi^2$ per degree of freedom, $\chi_{\rm DF}^2$.
These values and their combinations are also given in Table~\ref{t1}.

\begin{table}
\begin{center}
\caption{Values of the $J/\psi$ and $\psi^\prime$ matrix elements resulting
from the minimum-$\chi^2$ fits to the CDF data \protect\cite{abe}.
$M_r^\psi$ is defined in Eq.~(\ref{eq:mr}).}
\label{t1}
\smallskip
\begin{tabular}{|c|c|c|}
\hline\hline
$\psi$ & $J/\psi$ & $\psi^\prime$ \\
\hline
$\left\langle{\cal O}^\psi[\,\underline{1},{}^3\!S_1]\right\rangle$ &
$(7.63\pm0.54)\cdot10^{-1}$~GeV$^3$ & $(4.40\pm0.43)\cdot10^{-1}$~GeV$^3$ \\
$\left\langle{\cal O}^\psi[\,\underline{8},{}^3\!S_1]\right\rangle$ &
$(3.94\pm0.63)\cdot10^{-3}$~GeV$^3$ & $(6.20\pm0.95)\cdot10^{-3}$~GeV$^3$ \\
$M_r^\psi$ & 
$(6.52\pm0.67)\cdot10^{-2}$~GeV$^3$ & $(1.79\pm0.51)\cdot10^{-2}$~GeV$^3$ \\
$r$ & 3.47 & 2.56 \\
$\chi_{\rm DF}^2$ fus.\ & 5.97/10 & 1.00/4 \\
$\chi_{\rm DF}^2$ fra.\ & 1.53/2 & 0.03/2 \\
$\chi_{\rm DF}^2$ tot.\ & 7.49/12 & 1.03/6 \\
\hline\hline
\end{tabular}
\end{center}
\end{table}

As is well known \cite{cho}, prompt hadroproduction of $J/\psi$ and
$\psi^\prime$ mesons is rather insensitive to the individual values of
$\left\langle{\cal O}^\psi[\,\underline{8},{}^1\!S_0]\right\rangle$ and
$\left\langle{\cal O}^\psi[\,\underline{8},{}^3\!P_0]\right\rangle$ as long as
$M_r^\psi$ is kept fixed.
This is very different for their production from $B$ decay.
This circumstance may be exploited in order to separately fix
$\left\langle{\cal O}^\psi[\,\underline{8},{}^1\!S_0]\right\rangle$ and
$\left\langle{\cal O}^\psi[\,\underline{8},{}^3\!P_0]\right\rangle$ using the
CLEO data \cite{bal}.
We choose to do this by adjusting the NRQCD results for the $B\to\psi+X$ 
branching fractions so that they agree with the measured values.
The outcome and the $\chi_{\rm DF}^2$ values achieved are summarized in 
Table~\ref{t2}.
The fact that the CLEO data favour negative values of
$\left\langle{\cal O}^{J/\psi}[\,\underline{8},{}^3\!P_0]\right\rangle$ was 
already observed in Ref.~\cite{pal}.
The $\psi^\prime$ case was not considered there.
The resulting $B\to\psi+X$ differential branching fractions 
$(1/\Gamma_B)d\Gamma/dk_\psi^\prime$ for $\psi=J/\psi,\psi^\prime$ are
compared with the CLEO data in Figs.~\ref{f2}(a) and (b), respectively.

\begin{table}
\begin{center}
\caption{Decompositions of $M_r^\psi$ ($\psi=J/\psi,\psi^\prime$) from
Table~\ref{t1} into
$\left\langle{\cal O}^\psi[\,\underline{8},{}^1\!S_0]\right\rangle$ and
$\left\langle{\cal O}^\psi[\,\underline{8},{}^3\!P_0]\right\rangle$ obtained 
by requiring that the NRQCD values of $B(B\to\psi+X)$ for prompt $\psi$
production through $B$ decay agree with the CLEO data \protect\cite{bal}.
The resulting values of $\chi_{\rm DF}^2$ are also given.}
\label{t2}
\smallskip
\begin{tabular}{|c|c|c|}
\hline\hline
$\psi$ & $J/\psi$ & $\psi^\prime$ \\
\hline
$\left\langle{\cal O}^\psi[\,\underline{8},{}^1\!S_0]\right\rangle$ &
$1.45\cdot10^{-1}$~GeV$^3$ & $-9.66\cdot10^{-3}$~GeV$^3$ \\
$\left\langle{\cal O}^\psi[\,\underline{8},{}^3\!P_0]\right\rangle$ &
$-5.51\cdot10^{-2}$~GeV$^5$ & $2.58\cdot10^{-2}$~GeV$^5$ \\
$B(B\to\psi+X)$ (prompt) & 0.800\% & 0.340\% \\
$\chi_{\rm DF}^2$ \ & 103/20 & 16.5/8 \\
\hline\hline
\end{tabular}
\end{center}
\end{table}

An alternative criterion for determining
$\left\langle{\cal O}^\psi[\,\underline{8},{}^1\!S_0]\right\rangle$ and
$\left\langle{\cal O}^\psi[\,\underline{8},{}^3\!P_0]\right\rangle$ is to
minimize the $\chi^2$ values of the theoretical decay spectrum with the
experimental data.
This leads to the results shown in Table~\ref{t3}.
We observe that the $B\to J/\psi+X$ and $B\to\psi^\prime+X$ branching 
fractions are reduced by 15\% and 17\% relative to Table~\ref{t2}, while the
quality of the fits is only marginally improved.
From the sign flip of
$\left\langle{\cal O}^{\psi^\prime}[\,\underline{8},{}^1\!S_0]\right\rangle$
we conclude that this matrix element is not very well constrained from the
CLEO data.
On the other hand, its value in Table~\ref{t3} is very small.

We remark that our fits to the CLEO data have bad $\chi_{\rm DF}^2$ values.
From Figs.~\ref{f2}(a) and (b), we observe that the data both at small and
large $J/\psi$ momenta are not fitted well.
At large $J/\psi$ momentum, the spectrum is influenced by contributions from
the $K$, $K^*$, and higher $K^*$ resonances, which can only be included in an
average way by our approach.
We think that the exact shape of the $J/\psi$ spectrum is not relevant after 
the boost to the Tevatron laboratory frame.
Of importance, however, is the absolute normalization of the spectrum, since 
the final result is directly proportional to it.
Therefore, the fit procedure leading to Table~\ref{t2}, which is faithful in 
the experimental normalization, should be preferred, and we use the results of
Table~\ref{t2} in the remainder of this paper.

\begin{table}
\begin{center}
\caption{Same as in Table~\ref{t2}, but from minimum-$\chi^2$ fits to the CLEO
data \protect\cite{bal}.}
\label{t3}
\smallskip
\begin{tabular}{|c|c|c|}
\hline\hline
$\psi$ & $J/\psi$ & $\psi^\prime$ \\
\hline
$\left\langle{\cal O}^\psi[\,\underline{8},{}^1\!S_0]\right\rangle$ &
$2.23\cdot10^{-1}$~GeV$^3$ & $1.26\cdot10^{-3}$~GeV$^3$ \\
$\left\langle{\cal O}^\psi[\,\underline{8},{}^3\!P_0]\right\rangle$ &
$-1.09\cdot10^{-1}$~GeV$^5$ & $1.56\cdot10^{-2}$~GeV$^5$ \\
$B(B\to\psi+X)$ (prompt) & 0.684\% & 0.282\% \\
$\chi_{\rm DF}^2$ \ & 93.5/20 & 13.4/8 \\
\hline\hline
\end{tabular}
\end{center}
\end{table}

We now turn to the $B\to\chi_{cJ}+X$ decays.
The relevant matrix elements for their description in the NRQCD framework are
$\left\langle{\cal O}^{\chi_{c0}}[\,\underline{1},{}^3\!P_0]\right\rangle$ and
$\left\langle{\cal O}^{\chi_{c0}}[\,\underline{8},{}^3\!S_1]\right\rangle$.
For our analysis, we adopt the value
$\left\langle{\cal O}^{\chi_{c0}}[\,\underline{1},{}^3\!P_0]\right\rangle=
(8.80\pm2.13)\cdot10^{-2}$~GeV$^5$ from Ref.~\cite{gtb}, where is was
determined from the measured hadronic decay widths of the $\chi_{cJ}$ mesons.
Using the central value of this result, we then find
$\left\langle{\cal O}^{\chi_{c0}}[\,\underline{8},{}^3\!S_1]\right\rangle
=(1.39\pm0.17)\cdot10^{-3}$~GeV$^3$
from a LO fit to the CDF data on prompt $\chi_{cJ}$ hadroproduction \cite{abe}.
Using the NRQCD formalism described in Sect.~3 with this input, we evaluate
the $B\to\chi_{cJ}+X$ branching fractions for $J=0,1,2$ to be
$8.73\cdot10^{-5}$, $5.38\cdot10^{-4}$, and $3.90\cdot10^{-4}$, respectively.
These values are incompatible with the CLEO results,
$B(B\to\chi_{c1}+X)=(4.0\pm0.6\pm0.4)\cdot10^{-3}$ and
$B(B\to\chi_{c2}+X)=(2.5\pm1.0\pm0.3)\cdot10^{-3}$ \cite{bal}, where the first
error is statistical and the second systematic.
$B\to\chi_{c0}+X$ decays were not observed by CLEO.
Obviously, the CDF data on $p\bar p\to\chi_{cJ}+X$ and the CLEO data on
$B\to\chi_{cJ}+X$ cannot be simultaneously interpreted to LO in the NRQCD 
framework.
The inclusion of the NLO corrections to the $B\to\chi_{cJ}+X$ decay width does
not remove this discrepancy \cite{ben}.
In want of a theoretically satisfactory solution to this problem, we choose a
purely phenomenological description.
We calculate the momentum spectrum of the $J/\psi$ mesons from $B$ decay via
the $\chi_{cJ}$ states to LO in NRQCD, using the above matrix elements, an
average $\chi_{cJ}$-meson mass of 3.495~GeV, and the values of
$B(\chi_{cJ}\to J/\psi+\gamma)$ from Ref.~\cite{cas}, and adjust its
normalization in accordance with the CLEO result, to be 
$(1.3\pm0.2)\cdot10^{-3}$ \cite{bal}.
Without this normalization factor, we would have
$\sum_{J=0}^2B(B\to\chi_{cJ}+X)B(\chi_{cJ}\to J/\psi+\gamma)
=2.05\cdot10^{-4}$.
The CLEO Collaboration did not specify experimental data on the momentum
spectrum of the $J/\psi$ mesons originating from the $\chi_{cJ}$ states.
However, they published the measured decay spectrum for the combined $J/\psi$
sample, including prompt production and feed-down from $\chi_{c1}$,
$\chi_{c2}$, and $\psi^\prime$ states.
In Fig.~\ref{f2}(c), these data are compared with the superposition of our
NRQCD results.
Notice that both experimental and theoretical results have the same 
normalization.

\boldmath
\section{Charmonium production from $b$-hadron decay in $p\bar p$ collisions}
\unboldmath

In this section, we present our predictions for the inclusive production of
$J/\psi$ and $\psi^\prime$ mesons originating from $b$-hadron decay in
$p\bar p$ collisions at the Tevatron and compare them with available CDF
\cite{abe} and D0 \cite{abb} data.
As described in Sect.~3, the cross section of this process emerges from the 
one of inclusive $b$-hadron production by convolution with the
longitudinal-momentum distribution of the $J/\psi$ and $\psi^\prime$ mesons
from $b$-hadron decay appropriately boosted along the $b$-hadron flight
direction.
In Fig.~\ref{f3}, we investigate how the $B\to J/\psi+X$ branching fraction
$(1/\Gamma_B)d\Gamma(x,P_B)/dx$ differential in $x=k_L^\prime/P_B$, as given
by Eqs.~(\ref{eq:lon}) or (\ref{eq:alt}), is distorted by a boost to the frame
where the $B$ meson has nonvanishing three-momentum $P_B$.
Specifically, we consider the cases $P_B=5$, 10, and 20~GeV.
For comparison, also the asymptotic line shape according to 
Eq.~(\ref{eq:inf}), which refers to the infinite-momentum frame, is shown.
We observe that the finite-$P_B$ results rapidly converge towards the 
asymptotic form.
In order to obtain the effective $i\to J/\psi$ FF's via $B$ decay, we need to
convolute the $B\to J/\psi+X$ decay distribution shown in Fig.~\ref{f3} with
our $i\to B$ FF's \cite{bin} according to Eq.~(\ref{eq:eff}).

Having fixed all relevant input from fits to the OPAL data on $e^+e^-\to B+X$
\cite{ale}, the CDF data on $p\bar p\to\psi+X$ via prompt production 
\cite{abe}, and the CLEO data on $B\to\psi+X$ \cite{bal}, we are now in a
position to make absolute predictions for charmonium production via $b$-hadron
decay in $p\bar p$ collisions.
In Figs.~\ref{f4}(a) and (b), we consider $J/\psi$ production, summing over
the prompt channel and the feed-down channels via $\chi_{cJ}$ and
$\psi^\prime$ states.
In Fig.~\ref{f4}(a), the $p_T$ distribution $d\sigma/dp_T$ measured by CDF 
\cite{abe} in the central region of the detector, for $|y|<0.6$, is compared 
with our LO and NLO predictions based on the CTEQ4 \cite{lai} and MRST 
\cite{mar} PDF's and scale choice $\xi=1$.
In Fig.~\ref{f4}(b), the $\xi$ dependence of these predictions is analyzed for 
three representative values of $p_T$.
As a rule, the theoretical uncertainty may be estimated from three criteria:
(i) the difference between the LO and NLO predictions;
(ii) the shift due to scale variations;
and (iii) the variation between different PDF sets.
All these points may be studied with the help of Figs.~\ref{f4}(a) and (b).
We find good agreement with the experimental data, especially in the upper
$p_T$ range, where the theoretical uncertainty is smallest.
As expected, the NLO results are there more stable under scale variations than 
the LO ones.
At the lower end of the $p_T$ spectrum, the NLO results undershoot the data, 
but exhibit sizeable normalization uncertainties.
The CTEQ and MRST PDF's lead to very similar results, except at very low 
$p_T$.
In Fig.~\ref{f5}, the $J/\psi$ analysis of Fig.~\ref{f4}(a) is repeated for
$\psi^\prime$ mesons.
Here, the agreement between experimental data and NLO predictions is also good
at low $p_T$, where the theoretical uncertainty is largest.

The CDF \cite{abe} and D0 \cite{abb} Collaborations also published their full
data samples on $J/\psi$ production without discriminating between prompt
production, feed-down from charmonium states with higher masses, and 
$b$-hadron decay.
These cross sections are both differential in $p_T$, but are complementary in 
the sense that the CDF data are integrated over the central region, with
$|y|<0.6$, while the D0 data are sampled in the forward and backward
directions, with $2.5<|y|<3.7$.
In Figs.~\ref{f6}(a) and (b), we compare these data with our predictions 
obtained from the formalism explained here and in Ref.~\cite{bak} with the
nonperturbative charmonium matrix elements determined in Sect.~4.
We work at LO using the CTEQ4L PDF's \cite{lai}.
Apart from the total contributions, also the partial contributions due to
prompt production, feed-down from $\chi_{cJ}$ mesons, feed-down from
$\psi^\prime$ mesons, and $b$-hadron decay (summing over the prompt and
feed-down channels) are shown.
In the case of CDF, there is good agreement over the full $p_T$ range 
considered.
This is not surprising because the contributing matrix elements were actually
determined from the prompt $J/\psi$, $\chi_{cJ}$, and $\psi^\prime$ data
samples taken in the same experiment.
Furthermore, we already know from Fig.~\ref{f4}(a) that the predicted
$b$-hadron decay contribution agrees well with the CDF data.
So far, the D0 Collaboration did not present separate cross sections for the
individual channels of $J/\psi$ production.
The agreement between their combined cross section and the theoretical 
prediction, which is essentially tuned to fit the CDF data, is remarkably
good, except at the upper end of the $p_T$ spectrum.
In a way, Fig.~\ref{f6}(b) represents an indirect comparison between the CDF 
and D0 data on $J/\psi$ production, which refer to different kinematic 
regimes.
We notice that, at large $p_T$, the cross sections of the prompt and
$b$-hadron channels are of the same order for the CDF central production
[see Fig.~\ref{f6}(a)], whereas, in the case of D0 forward (backward)
production, the prompt channel plays the dominant r\^ole [see
Fig.~\ref{f6}(b)].
This means that the cross sections of these two channels have different
rapidity dependences.
This could be checked experimentally if these cross sections were measured 
separately also in the forward (backward) direction.

\section{Conclusions}

We considered the hadroproduction in $p\bar p$ collisions of $b$ hadrons which 
subsequently decay to charmonia and presented theoretical predictions for the
$p_T$ distribution of the latter.
The formation of the $b$ hadrons was described in the QCD-improved parton 
model with FF's fitted to $e^+e^-$ data on $B$-meson production \cite{bin}.
In want of detailed experimental information of the FF's of $B_s$ mesons and
$\Lambda_b$ baryons, we approximately accounted for their contributions by
appropriately adjusting the normalization of our $B$ FF's \cite{bin}.
The $B$ decays to the various charmonium states were treated in NRQCD
\cite{pal} with nonperturbative matrix elements determined from the CDF
\cite{abe} and CLEO \cite{bal} data on prompt-charmonium production in
$p\bar p$ collisions and $B$ decay, respectively.
In the case of the $B\to\chi_{cJ}+X$ decays, the NRQCD results for their
branching fractions came out considerably smaller than the CLEO results 
\cite{bal}, so that we had to include a phenomenological magnification factor.
Furthermore, we assumed that the decays of $B_s$ mesons and $\Lambda_b$
baryons to charmonia, for which no data exist so far, can be described on the
same footing as the $B$ decays.
We compared the CDF data on $J/\psi$ and $\psi^\prime$ production from 
$b$-hadron decay \cite{abe} with our theoretical predictions and found good
agreement.
We estimated the theoretical uncertainty by comparing LO and NLO predictions,
by scale variation, and by varying the PDF's and found that it was of order
$\pm25\%$ for $p_T\agt13$~GeV.
We also compared the full CDF \cite{abe} and D0 \cite{abb} data samples on
$J/\psi$ production without distinguishing between prompt, feed-down, and
$b$-hadron decay channels with our predictions.
Again, we found good agreement.
Since the CDF and D0 data refer to different ranges of rapidity, this may also
be regarded as an indirect comparison between them.

\bigskip
\centerline{\bf ACKNOWLEDGMENTS}
\smallskip\noindent
One of us (G.K.) is grateful to the Theory Group of the
Werner-Heisenberg-Institut for the hospitality extended to him during a visit
when this paper was prepared.

\newpage

\begin{figure}
\centerline{\bf FIGURE CAPTIONS}
\bigskip

\caption{\protect\label{f1} Fit to the CDF data on the inclusive
hadroproduction of prompt $\psi^\prime$ mesons \protect\cite{abe},
which come in the form of $d\sigma/dp_T$ integrated over $|y|<0.6$ as a
function of $p_T$.
Only the fusion results are shown.}

\caption{\protect\label{f2} Fit to the CLEO data on the production through
$B$ decay of (a) prompt $J/\psi$ mesons, (b) $\psi^\prime$ mesons, and (c)
$J/\psi$ mesons including those from the feed-down of $\chi_{c1}$,
$\chi_{c2}$, and $\psi^\prime$ mesons \protect\cite{bal}.}

\caption{\protect\label{f3} $B\to J/\psi+X$ (prompt) branching fraction
differential in the $J/\psi$ longitudinal momentum fraction relative to the
$B$ momentum $P_B$ for various values of $P_B$.
The result for $P_B=\infty$ is evaluated from Eq.~(\ref{eq:inf}).}

\caption{\protect\label{f4} The CDF data on the inclusive hadroproduction of
$b$ hadrons decaying to $J/\psi$ mesons \protect\cite{abe} are compared with LO
and NLO predictions evaluated with CTEQ4 \protect\cite{lai} and MRST
\protect\cite{mar} proton PDF's.
(a) $p_T$ dependence for scale choice $\xi=1$ and
(b) $\xi$ dependence for selected values of $p_T$.}

\caption{\protect\label{f5} The CDF data on the inclusive hadroproduction of
$b$ hadrons decaying to $\psi^\prime$ mesons \protect\cite{abe} are compared
with LO and NLO predictions evaluated with CTEQ4 \protect\cite{lai} and MRST
\protect\cite{mar} proton PDF's.}

\caption{\protect\label{f6} The (a) CDF \protect\cite{abe} and (b) D0
\protect\cite{abb} data on the inclusive hadroproduction of $J/\psi$ mesons
from all channels are compared with LO predictions evaluated with CTEQ4L
\protect\cite{lai} proton PDF's.}

\end{figure}

\newpage
\begin{figure}[ht]
\epsfig{figure=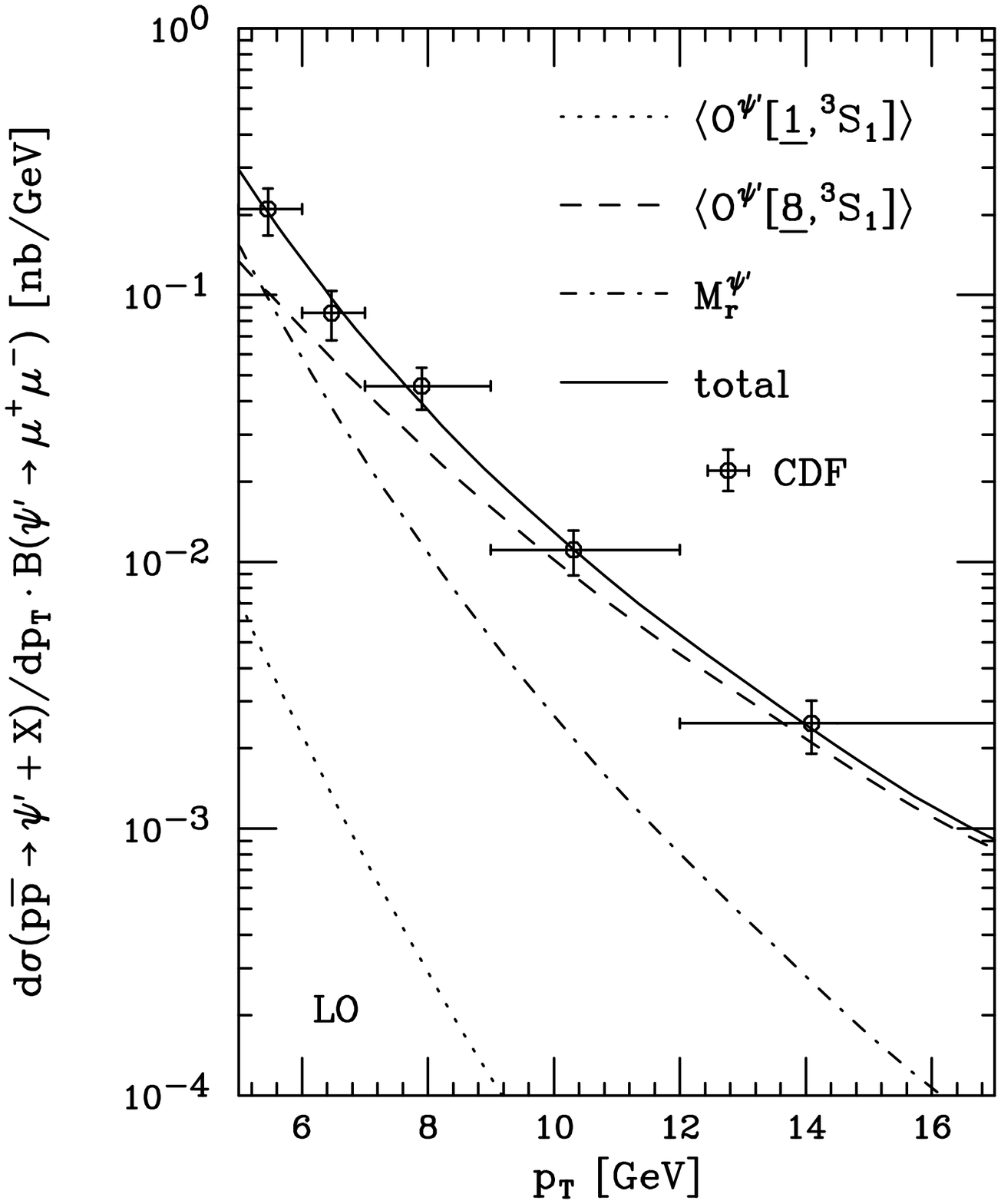,width=\textwidth}
\centerline{\Large\bf Fig.~1}
\end{figure}

\newpage
\begin{figure}[ht]
\epsfig{figure=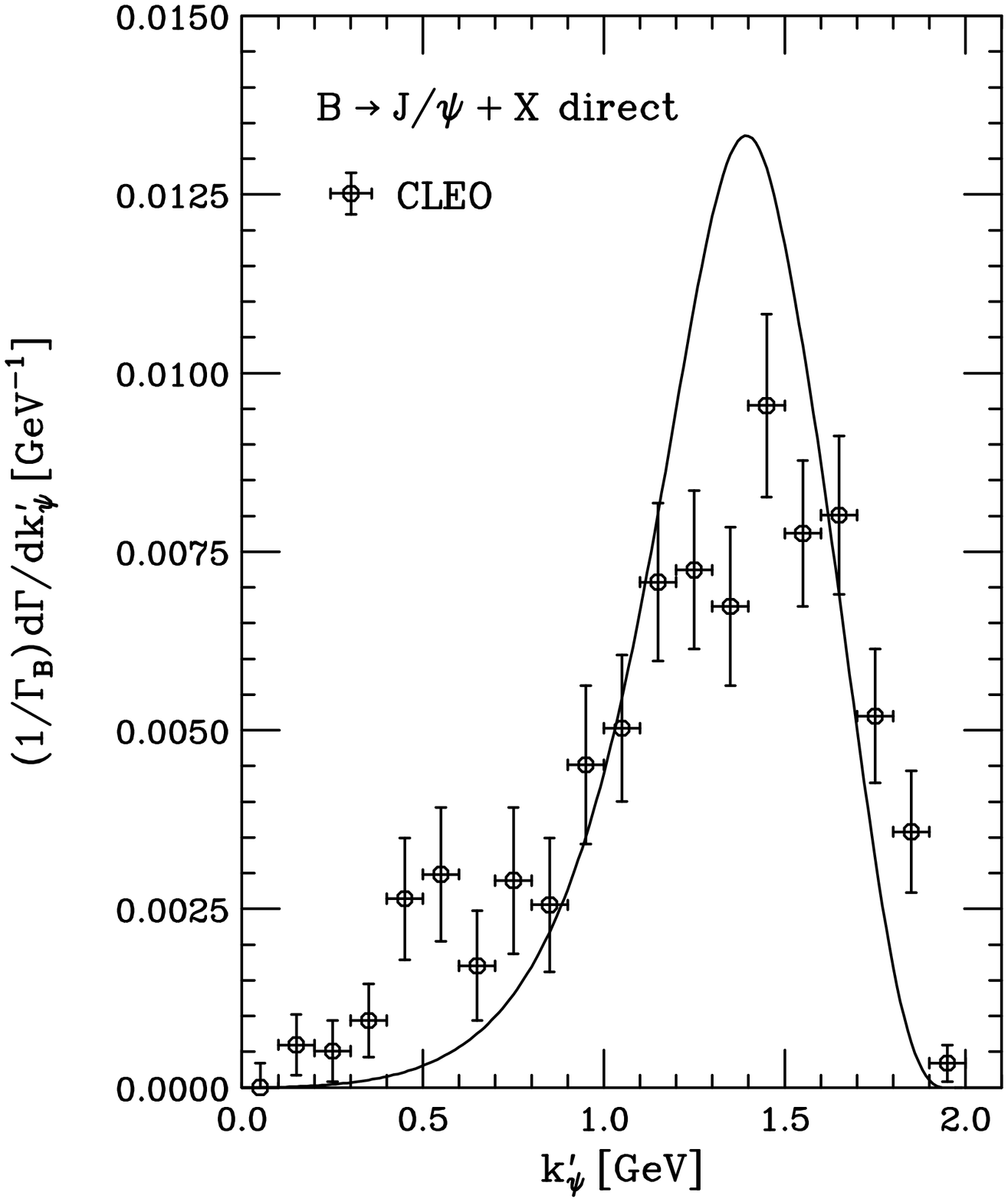,width=\textwidth}
\centerline{\Large\bf Fig.~2a}
\end{figure}

\newpage
\begin{figure}[ht]
\epsfig{figure=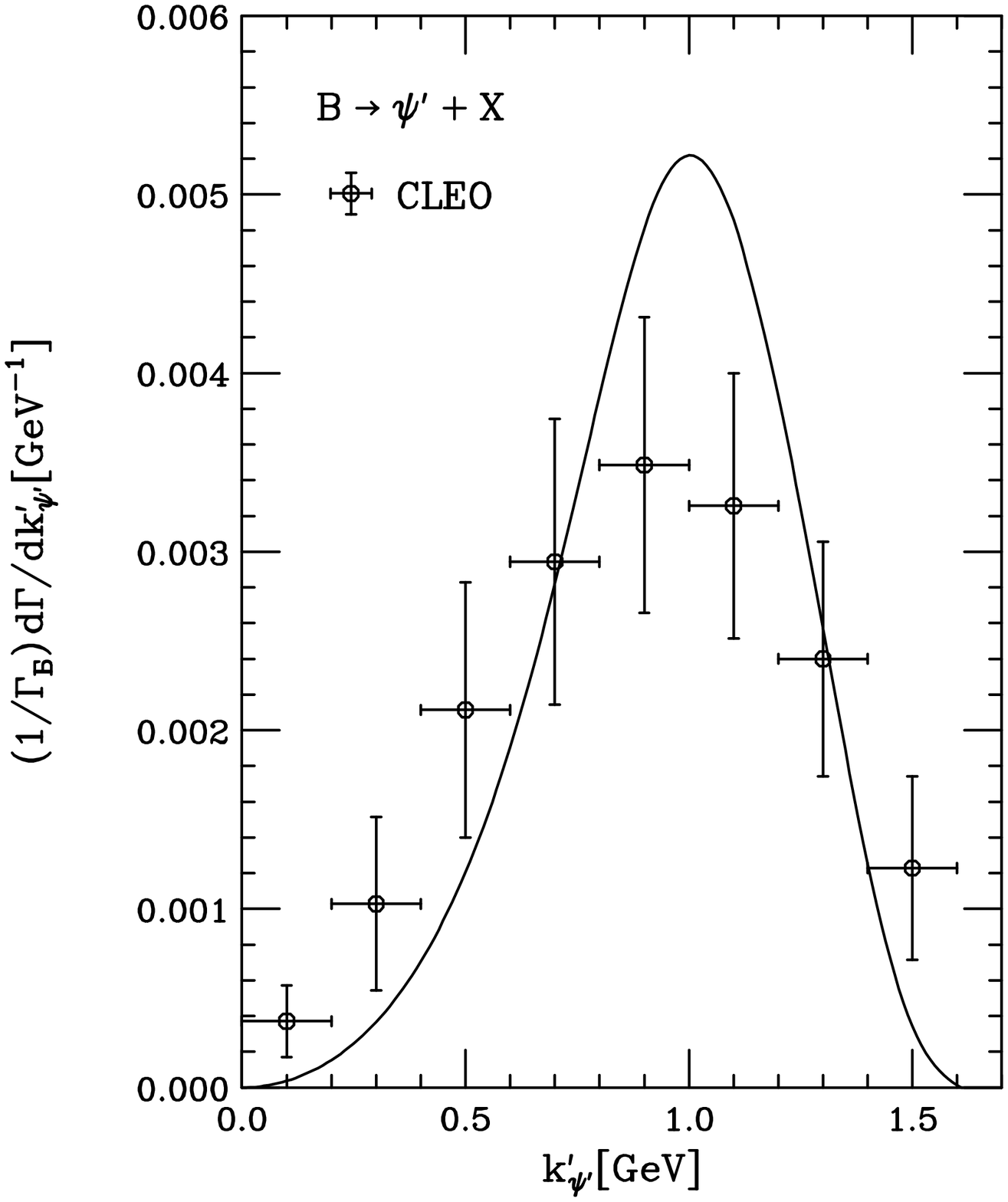,width=\textwidth}
\centerline{\Large\bf Fig.~2b}
\end{figure}

\newpage
\begin{figure}[ht]
\epsfig{figure=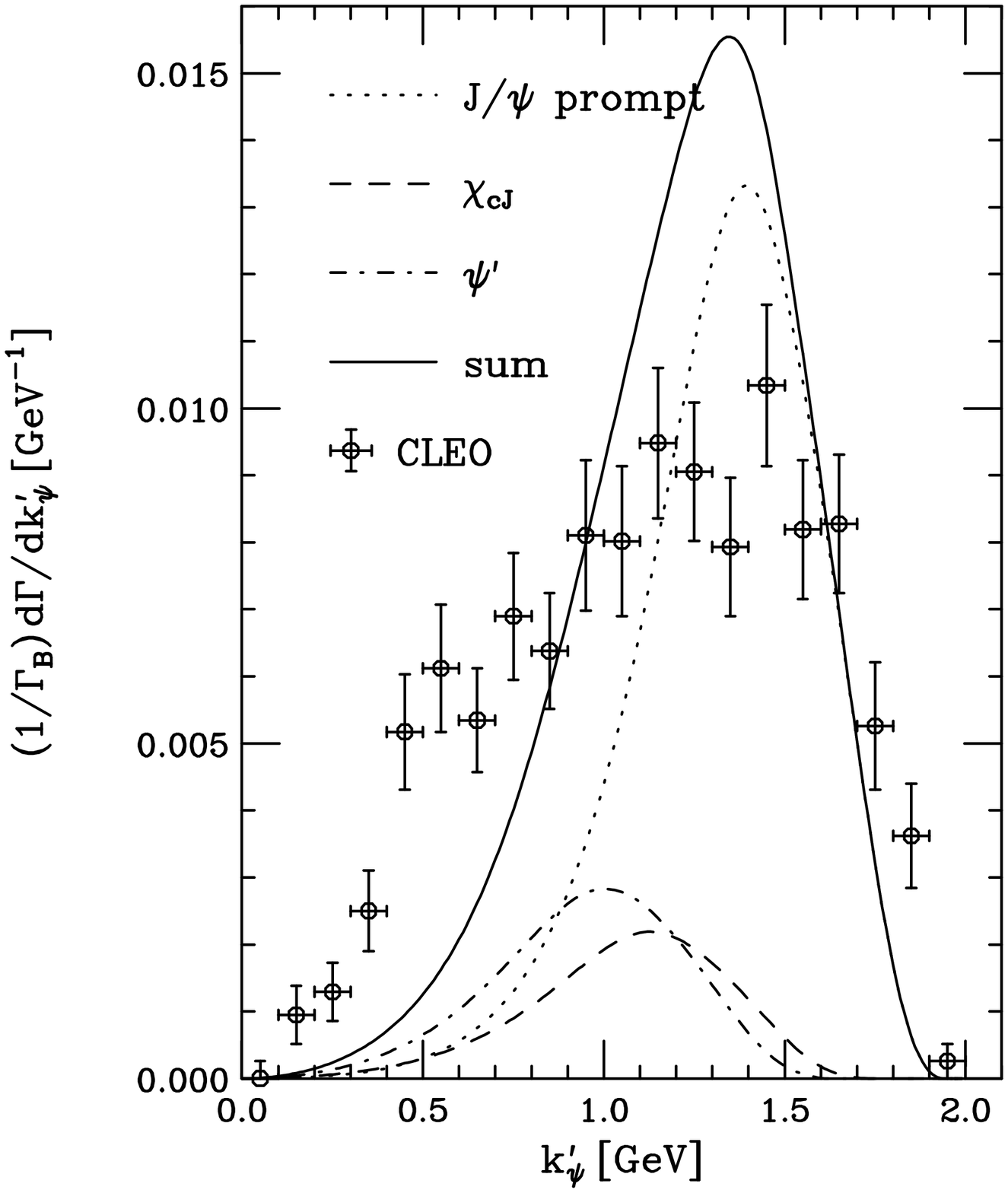,width=\textwidth}
\centerline{\Large\bf Fig.~2c}
\end{figure}

\newpage
\begin{figure}[ht]
\epsfig{figure=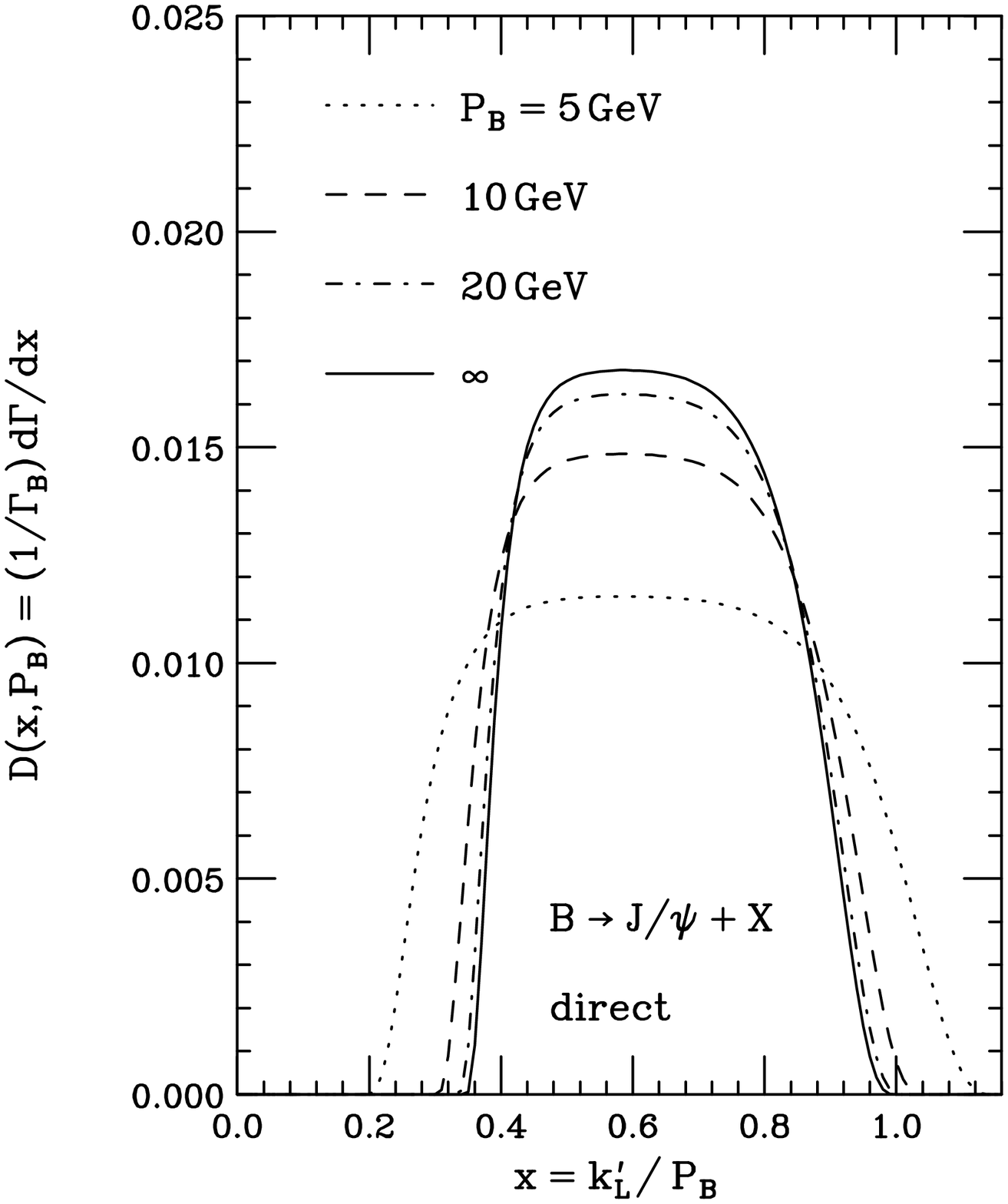,width=\textwidth}
\centerline{\Large\bf Fig.~3}
\end{figure}

\newpage
\begin{figure}[ht]
\epsfig{figure=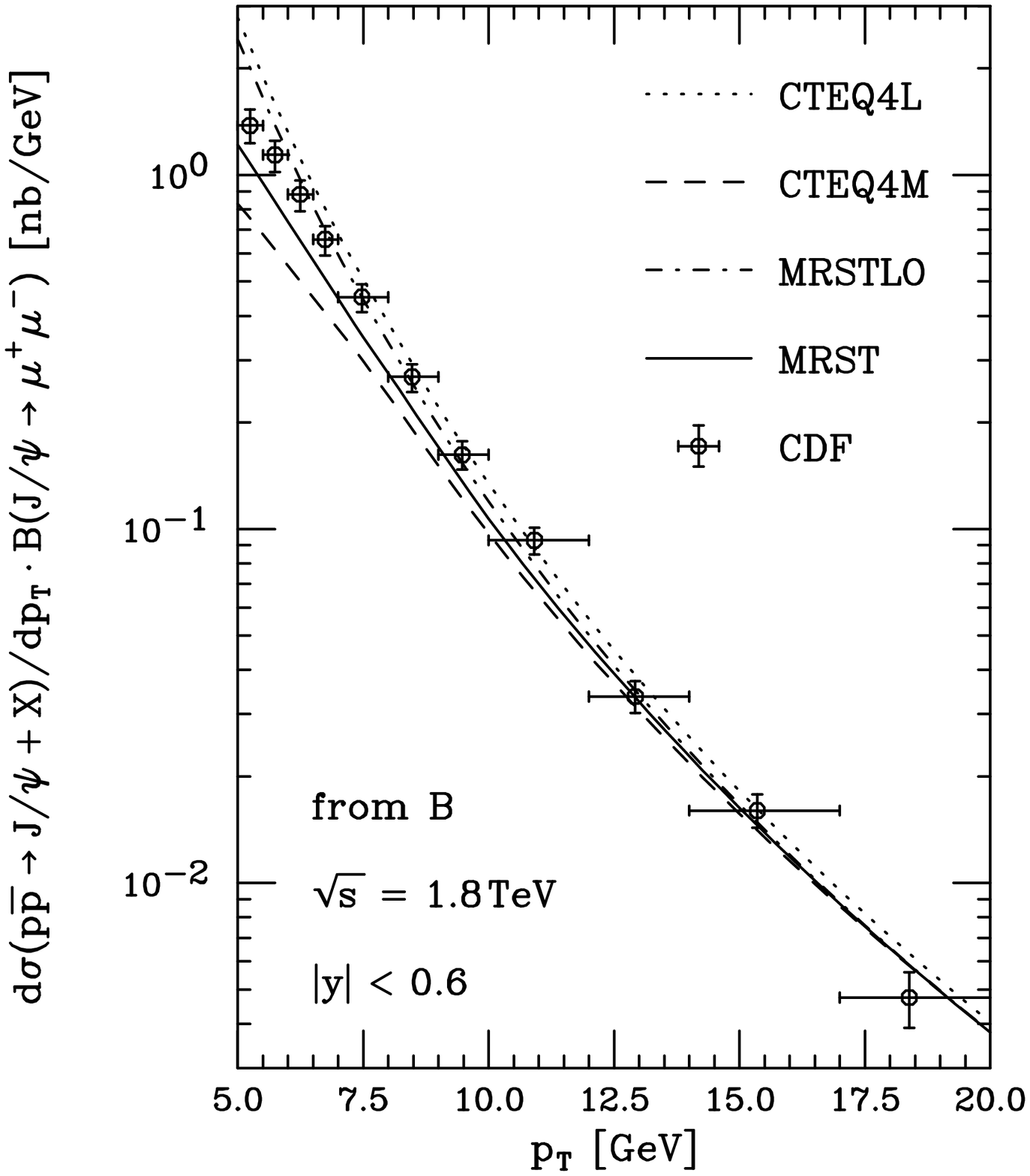,width=\textwidth}
\centerline{\Large\bf Fig.~4a}
\end{figure}

\newpage
\begin{figure}[ht]
\epsfig{figure=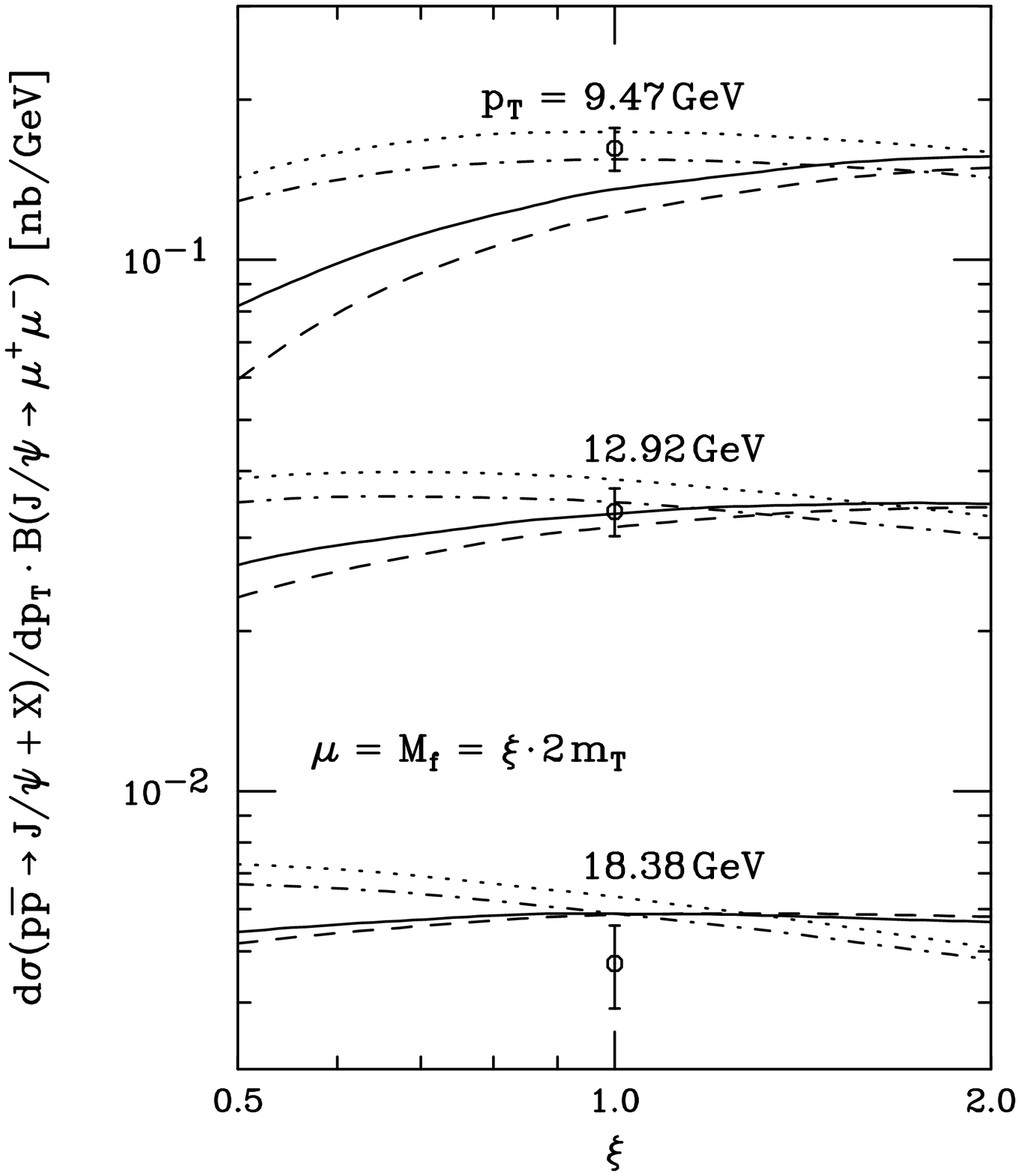,width=\textwidth}
\centerline{\Large\bf Fig.~4b}
\end{figure}

\newpage
\begin{figure}[ht]
\epsfig{figure=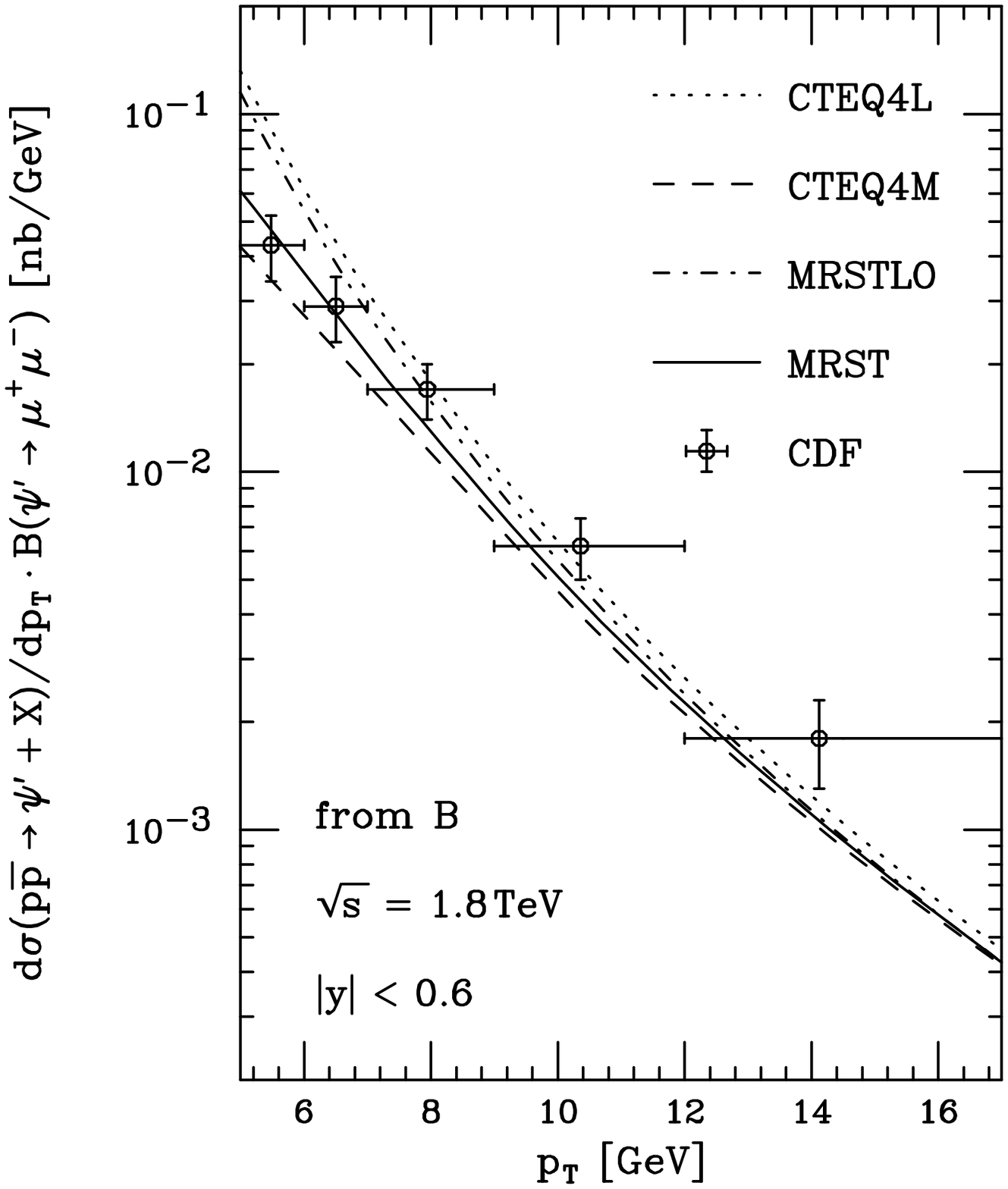,width=\textwidth}
\centerline{\Large\bf Fig.~5}
\end{figure}

\newpage
\begin{figure}[ht]
\epsfig{figure=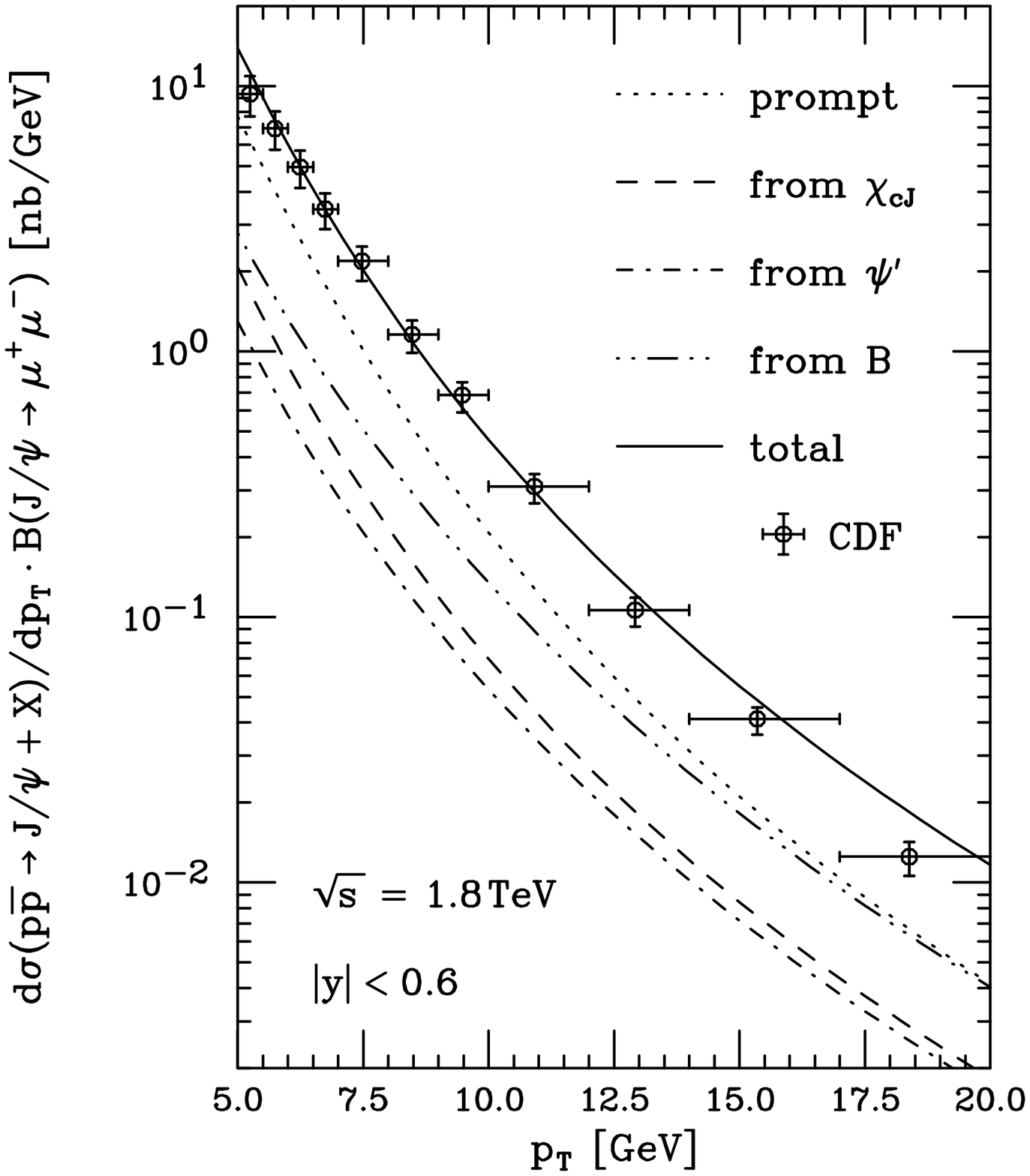,width=\textwidth}
\centerline{\Large\bf Fig.~6a}
\end{figure}

\newpage
\begin{figure}[ht]
\epsfig{figure=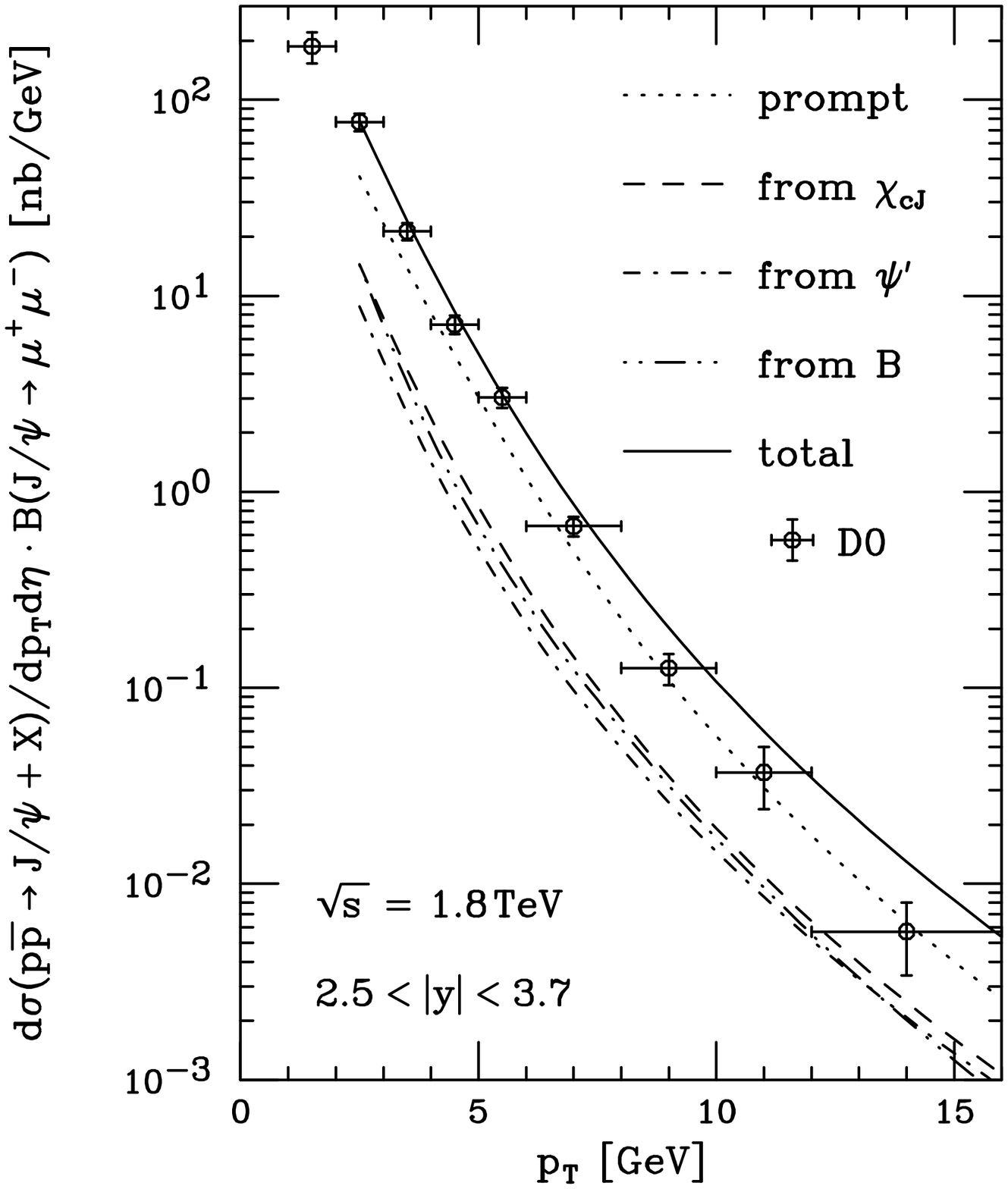,width=\textwidth}
\centerline{\Large\bf Fig.~6b}
\end{figure}


\begin{thebibliography}{99}
 
\bibitem{abe} CDF Collaboration, F. Abe {\it et al.},
Phys.\ Rev.\ Lett.\ {\bf79}, 572 (1997); {\bf79}, 578 (1997);
T. Daniels, Ph.~D. thesis, Massachusetts Institute of Technology (1997).

\bibitem{nas} P. Nason, S. Dawson, and R. K. Ellis,
Nucl.\ Phys.\ {\bf B303}, 607 (1988);
{\bf B327}, 49 (1989);
for a review, see S. Frixione, M. L. Mangano, P. Nason, and G. Ridolfi, in
{\it Heavy Flavors II}, edited by A. J. Buras and M. Lindner (World
Scientific, Singapore, 1998), p.~609.

\bibitem{bal} CLEO Collaboration, R. Balest {\it et al.},
Phys.\ Rev.\ D {\bf52}, 2661 (1995).

\bibitem{pet} C. Peterson, D. Schlatter, I. Schmitt, and P. M. Zerwas,
Phys.\ Rev.\ D {\bf27}, 105 (1983).

\bibitem{chr} J. Chrin,
Z. Phys.\ C {\bf36}, 163 (1987).

\bibitem{kni} B. A. Kniehl, M. Kr\"amer, G. Kramer, and M. Spira,
Phys.\ Lett.\ B {\bf356}, 539 (1995).

\bibitem{ale} OPAL Collaboration, G. Alexander {\it et al.},
Phys.\ Lett.\ B {\bf364}, 93 (1995).

\bibitem{bin} J. Binnewies, B. A. Kniehl, and G. Kramer,
Phys.\ Rev.\ D {\bf58}, 034016 (1998).

\bibitem{laa} CDF Collaboration, A. Laasanen {\it et al.},
Report No.\ FERMILAB-Conf-96/198-E (July 1996).

\bibitem{pal} W. F. Palmer, E. A. Paschos, and P. H. Soldan,
Phys.\ Rev.\ D {\bf56}, 5794 (1997).

\bibitem{bra} E. Braaten, K. Cheung, and T. C. Yuan,
Phys.\ Rev.\ D {\bf48}, R5049 (1993).

\bibitem{cha} C.-H. Chang and Y.-Q. Chen,
Phys.\ Rev.\ D {\bf49}, 3399 (1994).

\bibitem{cas} Particle Data Group, C. Caso {\it et al.},
Eur.\ Phys.\ J. C {\bf3}, 1 (1998).

\bibitem{bod} G. T. Bodwin, E. Braaten, and G. P. Lepage,
Phys.\ Rev.\ D {\bf51}, 1125 (1995); {\bf55}, 5853(E) (1997).

\bibitem{abb} D0 Collaboration, B. Abbott {\it et al.},
Phys.\ Rev.\ Lett.\ {\bf82}, 35 (1999).

\bibitem{bor} F. M. Borzumati, B. A. Kniehl, and G. Kramer,
Z. Phys.\ C {\bf57}, 595 (1993), and papers cited therein.

\bibitem{lai} H. L. Lai, J. Huston, S. Kuhlmann, F. Olness, J. Owens, D. Soper,
W. K. Tung, and H. Weerts,
Phys.\ Rev.\ D {\bf55}, 1280 (1997).

\bibitem{mar} A. D. Martin, R. G. Roberts, W. J. Stirling, and R. S. Thorne,
Eur.\ Phys.\ J. C {\bf4}, 463 (1998); Phys.\ Lett.\ B {\bf443}, 301 (1998).

\bibitem{jbi} J. Binnewies, B. A. Kniehl, and G. Kramer,
Z. Phys.\ C {\bf76}, 677 (1997).

\bibitem{ali} A. Ali and E. Pietarinen,
Nucl.\ Phys.\ {\bf B154}, 519 (1979);
G. Altarelli, N. Cabibbo, G. Corbo, L. Maiani, and G. Martinelli,
Nucl.\ Phys.\ {\bf B208}, 365 (1982).

\bibitem{bak} B. A. Kniehl and G. Kramer,
Eur.\ Phys.\ J. C {\bf6}, 493 (1999).

\bibitem{cho} P. Cho and A. K. Leibovich,
Phys.\ Rev.\ D {\bf53}, 150 (1996); {\bf53}, 6203 (1996).

\bibitem{gtb} G. T. Bodwin, E. Braaten, and G. P. Lepage,
Phys.\ Rev.\ D {\bf46}, R1914 (1992).

\bibitem{ben} M. Beneke, F. Maltoni, and I.Z. Rothstein,
Report No.\ CERN-TH/98-240, CMU-9805, and hep-ph/9808360 (August 1998).

\end{thebibliography}
\end{document}